\documentclass[journal]{IEEEtran}
\usepackage{subcaption} 

\ifCLASSINFOpdf
\else
   \usepackage[dvips]{graphicx}
\fi
\usepackage{url}

\usepackage{cite}
\usepackage{hyperref}
\usepackage{graphicx}
\usepackage{color}
\usepackage{amsfonts,amssymb}
\usepackage[linesnumbered,ruled,vlined]{algorithm2e}
\usepackage{amsmath}
\usepackage{threeparttable}
\usepackage{multirow}
\usepackage{booktabs} % 用于表格中线条控制
\usepackage{array} % 用于表格中列宽控制
\usepackage{makecell} % 用于表格中换行
\usepackage{float} % 用于控制图表位置

\begin{document}

\title{Radio-FM: A Foundation Model for Radio Signal Representation Learning and Its Applications}

\author{
    \vskip 1em
    Jinchao~Zhou,
    Wupeng~Xie,
    Zhuangzhi~Chen, \emph{Member,~IEEE},
    Yao~Lu,
    Qi~Xuan, \emph{Senior Member,~IEEE},
    Yun~Lin, \emph{Senior Member,~IEEE},
    and Guan~Gui, \emph{Fellow,~IEEE}
    \thanks{This work was supported in part by the National Natural Science Foundation of China under Grant 62301492. (\emph{Corresponding authors: Zhuangzhi Chen}.)}
    \thanks{Jinchao Zhou, Zhuangzhi Chen, Yao Lu and Qi Xuan are with the Institute of Cyberspace Security, Zhejiang University of Technology, Hangzhou 310023, China; Jinchao Zhou, Zhuangzhi Chen and  Qi Xuan are also with the Binjiang Institute of Artificial Intelligence, ZJUT, Hangzhou 310056, China (e-mail: jinchaozhou73@gmail.com; zzch@zjut.edu.cn; yaolu.zjut@gmail.com; xuanqi@zjut.edu.cn).}
    \thanks{Wupeng Xie is with Artificial Intelligence Institute of China Electronics Technology Group Corporation, Beijing 100041, China (e-mail: xiewupeng15@mails.ucas.ac.cn)}
    \thanks{Yun Lin is with the College of Information and Communication Engineering, Harbin Engineering University, Harbin 150000, China (email: linyun@hrbeu.edu.cn).}
    \thanks{Guan Gui is with the College of Telecommunications and Information Engineering, Nanjing University of Posts and Telecommunications, Nanjing 210003, China (email: guiguan@njupt.edu.cn).}
}

\markboth{}
{Shell \MakeLowercase{\textit{et al.}}: Bare Demo of IEEEtran.cls for IEEE Journals}
\maketitle

\begin{abstract}
Applying foundation models to the radio frequency (RF) domain presents unique challenges due to the intrinsic physical complexity of raw I/Q signals and the extreme heterogeneity of spectral data.
In this paper, we present Radio-FM, a scalable family of foundation models designed for universal radio signal representation learning.
Unlike standard architectures, Radio-FM employs dual-channel processing specifically optimized for I/Q independence while capturing cross-channel interactions through a lightweight attention mechanism.
To scale pretraining across heterogeneous multi-source corpora with highly variable sequence lengths, we propose a token-budgeted dynamic batching strategy coupled with channel-independent masked reconstruction.
We pretrain Radio-FM on a diverse collection of 15 datasets spanning modulation, radar, and communication domains, and rigorously evaluate it on 15 downstream benchmarks.
Experimental results show Radio-FM achieves state-of-the-art performance on 13 of 15 benchmarks, consistently improving across modulation, radar, emitter identification, wireless technology recognition, and wireless interference identification. 
Notably, it exhibits superior few-shot transferability, significantly outperforming existing baselines in data-scarce regimes, validating its potential as a general-purpose backbone for radio signal understanding.
\end{abstract}


\IEEEpeerreviewmaketitle

\section{Introduction\label{introduction}}

\begin{figure}[t]
    \centering
    \includegraphics[width=1.00\linewidth]{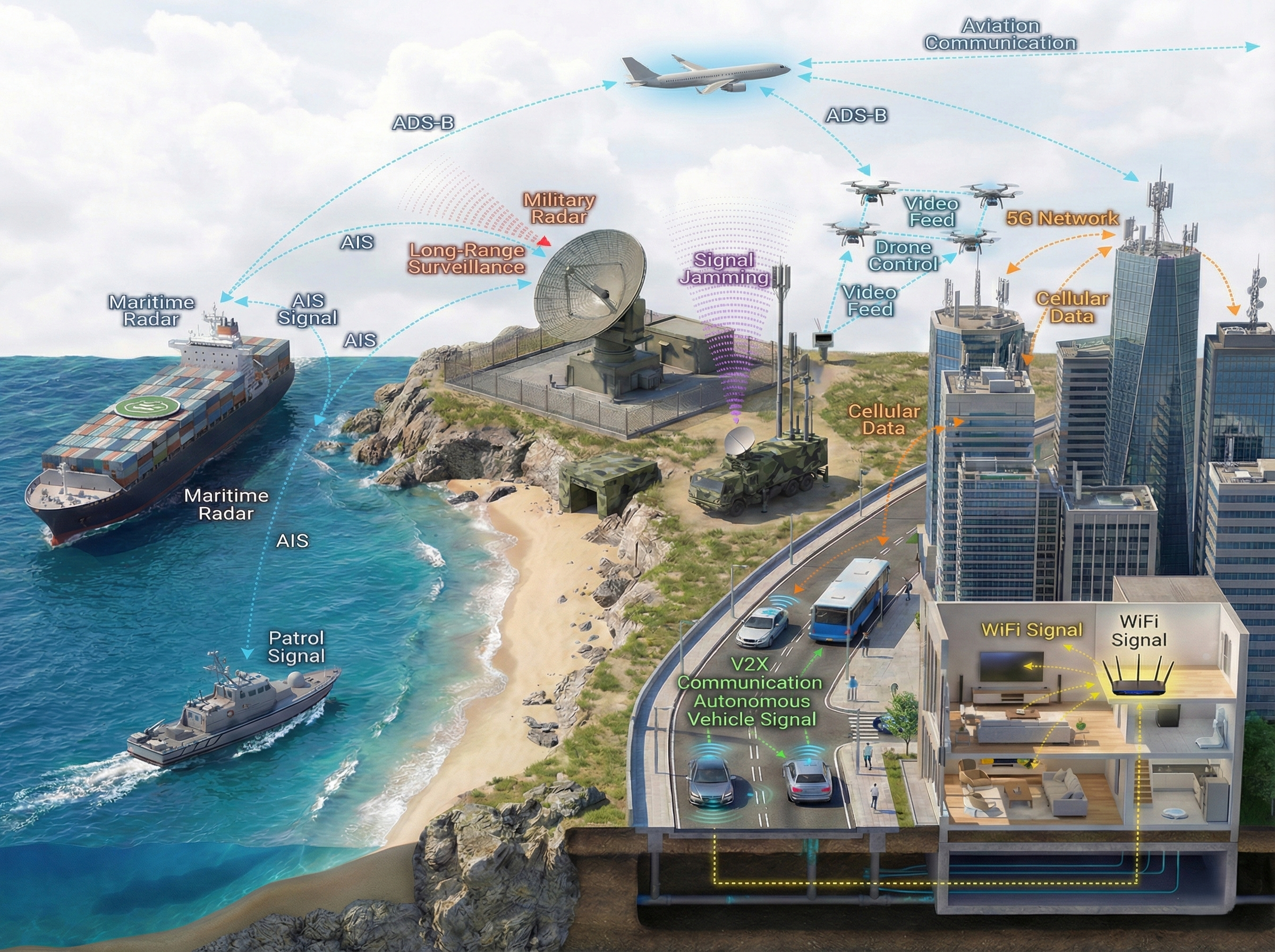}
    \caption{A representative electromagnetic ecosystem with heterogeneous signal sources and application scenarios. (AI-generated image using Nano Banana)}
    \label{fig:em_ecosystem}
\end{figure}

\IEEEPARstart{W}{ith} the rise of large-scale pretraining, the foundation-model paradigm has reshaped modern artificial intelligence, from natural language processing to computer vision.
By distilling transferable representations from massive unlabeled data, this paradigm offers a practical route to improving generalization across downstream tasks, which is equally critical for intelligent radio signal understanding, including spectrum monitoring, automatic modulation recognition (AMR), radar waveform recognition (RWC), specific emitter identification (SEI), wireless technology recognition (WTR), wireless interference identification (WII), and related tasks.
Fig.~\ref{fig:em_ecosystem} depicts a representative electromagnetic ecosystem with diverse emitters and acquisition conditions across sea, land, and air, motivating large-scale pretraining on heterogeneous radio I/Q signals for broad downstream adaptation.

Nevertheless, most radio signal understanding pipelines still rely on task-specific supervised learning, which faces three fundamental obstacles.
First, label scarcity: unlike natural images, acquiring high-quality labels for radio signals (e.g., emitter types and interference parameters) requires expensive equipment and expert annotation.
Second, limited generalization: wireless channels are highly dynamic and non-stationary; models trained on narrow or synthetic distributions often suffer severe performance degradation under real-world domain shifts.
Third, deployment fragmentation: the prevailing ``one-task-one-model'' approach stacks multiple dedicated models on resource-constrained edge devices, hindering scalable deployment.

To address these challenges, the community is increasingly turning to Foundation Models (FMs) that learn transferable representations via self-supervised pretraining on large-scale unlabeled data.
Motivated by successes in vision and language, recent efforts have explored pretraining across wireless modalities and deployment settings, including spectrogram-based modeling~\cite{aboulfotouh2024building,aboulfotouh2025wavesfm}, channel/CSI-centric pretraining~\cite{jiang2025towards,liu2025wifo}, and efficient on-device or federated deployment~\cite{hallaq2025tiny}; others further extend FMs to sensing/localization and generation~\cite{jiang2025scale,aboulfotouh2025multimodal,pan2025lwlm,chi2024rf,yang2025wirelessgpt}.
Meanwhile, research on raw I/Q signals, the native physical-layer representation, has advanced with MAE-style masked reconstruction (e.g., RIS-MAE~\cite{liu2025ris}) and spectrum foundation models (e.g., SpectrumFM~\cite{zhou2025spectrumfm} and EMind~\cite{luo2025emind}).
Despite this progress, existing approaches often struggle to satisfy three requirements simultaneously: (i) physical alignment, as treating I/Q signals as images or simple sequences can obscure their phase and amplitude structure; (ii) temporal consistency, as standard masking may break the positional correspondence needed for precise reconstruction; and (iii) training efficiency, as mixing heterogeneous datasets with widely varying lengths can cause severe computational load imbalance.

In this paper, we propose Radio-FM, a scalable family of foundation models for radio signal representation learning and its applications.
Radio-FM adopts a dual-channel processing backbone specifically optimized for I/Q independence that models intra-channel temporal dependencies and inter-channel correlations via a dual-level attention design. For self-supervised pretraining, we introduce channel-independent random masking and a position-preserving strategy that keeps the original patch indices during masking, enabling consistent alignment for masked reconstruction with Rotary Positional Embeddings (RoPE).
For efficient large-scale training, we further develop a multi-dataset pretraining recipe with a token/patch-budgeted dynamic batching mechanism, dataset-level sampling, and explicit distributed sharding to improve throughput, stability, and reproducibility on heterogeneous datasets.

We validate Radio-FM on a comprehensive benchmark suite. The pretraining stage uses 15 datasets spanning modulation, radar, and communication signals, and downstream evaluation covers 15 datasets across AMR, RWC, SEI, WTR, and WII tasks (see Table~\ref{tab:pre_training_datasets} and Table~\ref{tab:downstream_datasets}).

The main contributions are summarized as follows:
\begin{itemize}
    \item We propose Radio-FM, a general-purpose backbone featuring dual-channel processing specifically optimized for I/Q independence. Unlike standard architectures, it employs channel-independent processing and lightweight attention to jointly model signal dependencies.
    \item We develop a scalable pretraining framework tailored for heterogeneous radio spectra. We introduce channel-independent masked reconstruction to enhance representation learning, supported by an efficient token-budgeted dynamic batching mechanism that maximizes computational utilization.
    \item We conduct extensive evaluations on 15 diverse downstream benchmarks. Radio-FM achieves state-of-the-art performance on 12 tasks, outperforms baselines by over 20\% in few-shot regimes, and demonstrates clear scaling behavior from Tiny to XLarge variants.
\end{itemize}

The remainder of this paper is organized as follows. Section~\ref{sec_1} reviews related works. Section~\ref{sec_2} presents the Radio-FM framework and pretraining objective. Section~\ref{sec_3} reports experimental results, comparisons, and ablations. Finally, Section~\ref{sec_4} concludes the paper.

\section{Related Works\label{sec_1}}
The transition from task-specific supervised learning to general-purpose foundation models marks a pivotal shift in wireless AI. This section reviews the limitations of traditional paradigms and the emerging landscape of wireless foundation models.

\begin{figure*}[t]
    \centering
    \includegraphics[width=0.98\textwidth]{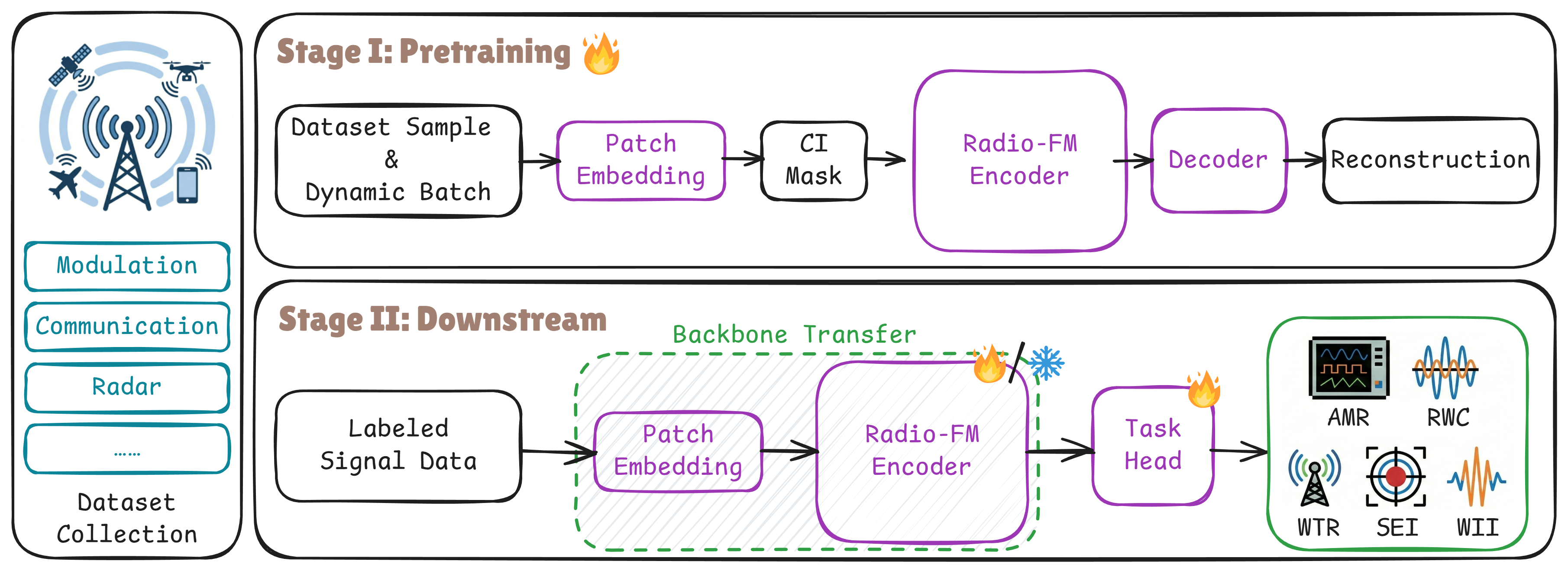}
    \caption{Overall training pipeline of Radio-FM, comprising three phases: (1) \textbf{Dataset Collection and Standardization}, where heterogeneous radio signals are unified via standardized preprocessing and efficient dynamic batching; (2) \textbf{Stage I: Pretraining}, where the encoder is trained on large-scale unlabeled data via masked reconstruction; and (3) \textbf{Stage II: Downstream Adaptation}, where the pretrained encoder is transferred to diverse downstream tasks via linear probing or fine-tuning.}
    \label{fig:pipeline}
\end{figure*}

\subsection{Deep Learning for Radio Signals and Limitations}
In cognitive radio and spectrum intelligence, the diversity of physical-layer objectives has fostered a broad landscape of task-tailored neural architectures~\cite{ya2022large}.
Representative examples span AMR~\cite{o2016convolutional,Tekbiyik2020robust,chen2021signet}, where models range from CNN/attention backbones on raw I/Q to constellation/time--frequency views, complex-valued processing, semi-supervision, and graph encoders~\cite{lin2020contour,tu2020complex,tu2018semi,xuan2022avgnet,yin2024dtsg};
RWC~\cite{10149618,huang2023multi}, where sequence and time--frequency encoders model waveform structure and dynamics;
SEI~\cite{9721895,reus2020trust,yan2025radio,wang2025tfmix}, where feature extractors and metric learning capture subtle hardware fingerprints under channel variation;
WTR~\cite{fontaine2019towards,fontaine2020multi}, where lightweight classifiers distinguish coexisting technologies across bands and environments;
and WII~\cite{schmidt2017wireless,zhang2019deep,luo2025emind}, where interference detectors identify heterogeneous interference patterns for spectrum access and mitigation.

However, most of these systems remain task-specific and label-driven, and they are often brittle under device, channel, and environment shifts.
In practice, extending to new tasks or operating conditions typically requires costly re-labeling and re-training, while maintaining multiple specialized models remains inefficient.
These limitations have catalyzed a shift towards Foundation Models (FMs) that learn transferable representations from large-scale unlabeled I/Q data.

\subsection{Wireless Foundation Models}
Inspired by the success of BERT and GPT, recent research has pivoted towards self-supervised learning (SSL) for wireless signals.
Existing foundation models differ primarily in their input modality (e.g., raw baseband I/Q, CSI, spectrograms, or multimodal combinations), which in turn shapes architectural choices and pretraining objectives.
Since our goal is to learn general-purpose representations from the native physical-layer measurement, we focus on raw I/Q-based foundation models and only briefly summarize other modalities.

Among I/Q-based FMs, EMind~\cite{luo2025emind} introduces an electromagnetic-signal foundation model trained on a large standardized dataset with masked reconstruction; it combines an encoder--decoder Transformer over tokenized I/Q streams with length-adaptive multi-signal packing and a hardware-aware training strategy to improve efficiency and transfer across heterogeneous tasks.
SpectrumFM~\cite{zhou2025spectrumfm} targets spectrum management with an encoder that integrates convolutional feature extraction and multi-head self-attention; it is pretrained on large-scale I/Q data via masked reconstruction and next-slot signal prediction, and adopts parameter-efficient fine-tuning for downstream tasks (e.g., AMC, WTR, spectrum sensing, and anomaly detection).
IQFM~\cite{mashaal2025iqfm} proposes a lightweight raw-I/Q foundation model pretrained via contrastive SSL on over-the-air multi-antenna I/Q; it leverages task-aware augmentations and supports efficient adaptation (e.g., with LoRA) to diverse tasks such as modulation classification, AoA estimation, beam prediction, and RF fingerprinting.
Beyond single-modality raw I/Q, other wireless FMs either broaden the set of supported modalities or leverage complementary channel representations. Multimodal WFM~\cite{aboulfotouh2025multimodal} jointly processes raw I/Q streams and image-like wireless modalities (e.g., spectrograms and CSI) and learns a shared representation via multimodal masked modeling, while WTR-FM~\cite{cheraghinia2025foundation} pretrains a Transformer on large-scale unlabeled acquired I/Q and CIR time-series for wireless technology recognition and localization. SkyLLM~\cite{SkyLLM2026Zhang} extracts raw I/Q samples of UAV RF signals, maps integrated I/Q, amplitude-phase and temporal features into the token embedding space of a frozen large language model (LLM) backbone, and realizes open-world UAV RF identification as well as unknown signal rejection.
Compared with prior I/Q-based FMs, Radio-FM emphasizes dual-channel processing specifically optimized for I/Q independence and scalable pretraining on heterogeneous datasets to support broad cross-task transfer.

\section{Method\label{sec_2}}

\begin{figure*}[t]
    \centering
    \includegraphics[width=0.95\textwidth]{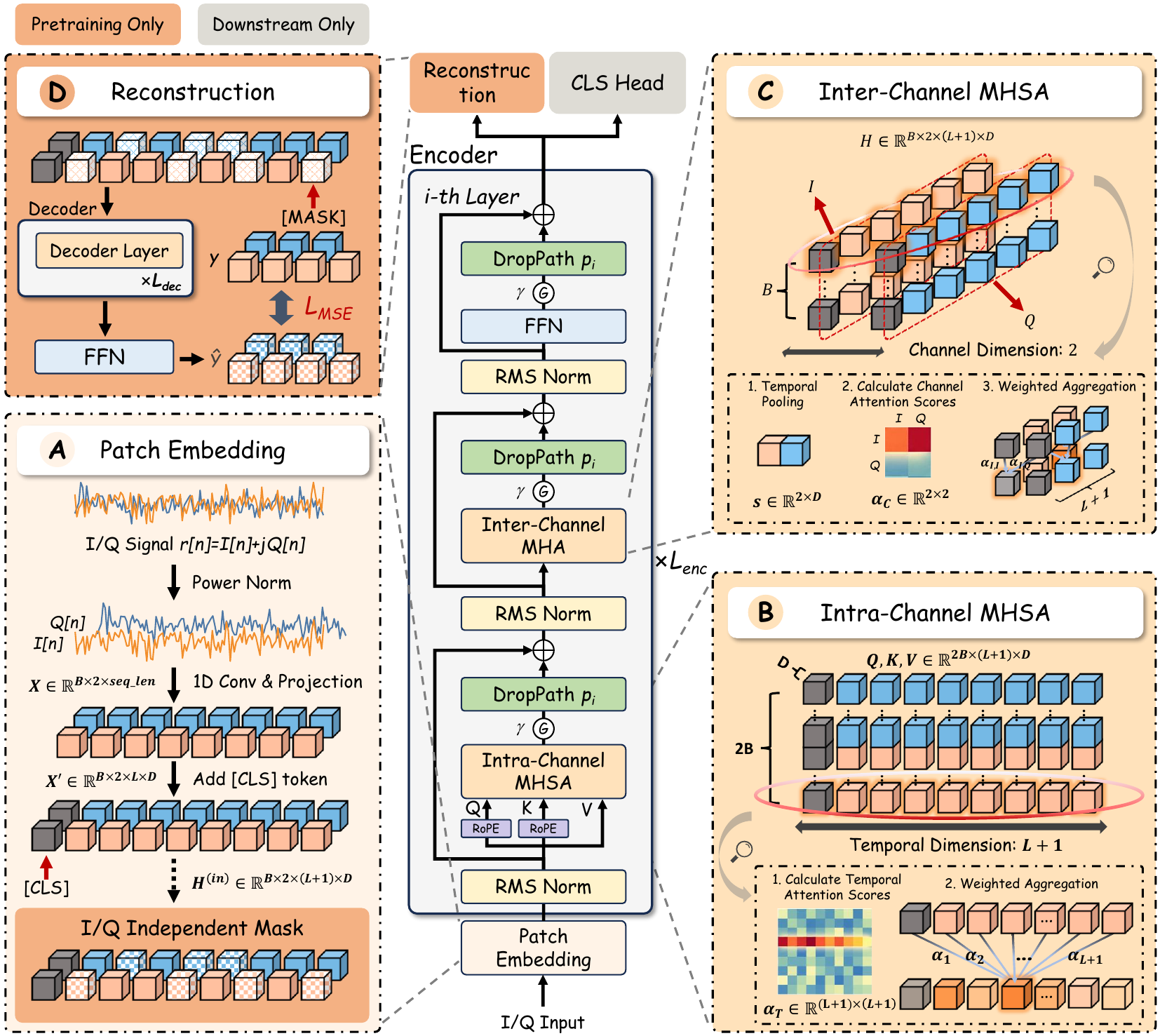}
    \caption{Model architecture of Radio-FM. The encoder performs intra-channel self-attention and inter-channel interaction, and is trained with channel-independent masked reconstruction in pretraining.}
    \label{fig:architecture}
\end{figure*}

This section presents Radio-FM, a two-stage framework for representation learning from raw radio I/Q signals. Fig.~\ref{fig:pipeline} summarizes the overall training pipeline, while Fig.~\ref{fig:architecture} illustrates the model architecture.

\subsection{Problem Statement and Training Paradigm}
In wireless communications and cognitive radio, a discrete-time complex baseband signal is commonly modeled as $r[n]=I[n]+jQ[n]$, where $I[n]$ and $Q[n]$ are the in-phase and quadrature components, and an instance consists of $l$ discrete-time samples indexed by $n=1,\ldots,l$.
After unified standardization, heterogeneous radio datasets can be represented as $\{\mathcal{D}_k\}_{k=1}^{K}$, where each $\mathcal{D}_k=\{r_i^{(k)}[n]\}_{i=1}^{N_k}$ comprises variable-length baseband sequences collected under diverse acquisition conditions.

Following the widely adopted two-stage paradigm for self-supervised representation learning (Fig.~\ref{fig:pipeline}), an encoder $f_{\theta}$ is first pretrained on large-scale unlabeled tokens $x$ via masked reconstruction. Let $\mathcal{M}(\cdot)$ denote a masking operator and $g_{\psi}$ a lightweight decoder; we optimize
\begin{equation}
\min_{\theta,\psi}\; \mathbb{E}_{x}\Big[\ell\big(x,\; g_{\psi}(f_{\theta}(\mathcal{M}(x)))\big)\Big],
\end{equation}
where $\ell(\cdot,\cdot)$ is a reconstruction loss. To improve computational utilization under severe length heterogeneity, dynamic batching is used with a maximum-token budget $C_{\max}$, forming batches $\mathcal{B}$ such that $\sum_{x\in\mathcal{B}} |x| \le C_{\max}$, where $|x|$ denotes the token length.

In the second stage, for each downstream task $t \in \mathcal{T}$ with labeled data $\{(x_i,y_i)\}$, a task head $h_{\phi}^{(t)}$ is attached and we minimize
\begin{equation}
\min_{\phi}\; \mathbb{E}_{(x,y)}\Big[\mathcal{L}^{(t)}\big(h_{\phi}^{(t)}(f_{\theta}(x)),\, y\big)\Big],
\end{equation}
either with a frozen encoder (linear probing) or by fine-tuning $\theta$ end-to-end.

\subsection{Input Processing and Patch Embedding}
Before being fed into the backbone network, the raw complex baseband waveform is transformed into the token sequence $x$ used for pretraining and downstream fine-tuning.
Given an input sequence of length $N$ (extracted from the raw instance of length $l$), denoted as $r[1{:}N]=I[1{:}N]+jQ[1{:}N]$, power normalization is first applied to standardize the amplitude scale.
Specifically, each sequence is normalized by its root-mean-square (RMS) power, $\sigma=\sqrt{\frac{1}{N}\sum_{n=1}^{N}|r[n]|^2}$, yielding $r'[n]=r[n]/\sigma$.
To preserve the I/Q structure while capturing local temporal patterns, we tokenize the in-phase and quadrature components independently using a 1D convolution with kernel size $P$ and stride $S$.
The resulting number of patches per channel can be computed as
\begin{equation}
L = \left\lfloor \frac{N - P}{S} \right\rfloor + 1
\end{equation}
Each patch is linearly projected to a $D$-dimensional embedding, and a learnable $\text{[CLS]}$ token is prepended to each channel sequence. The resulting input is $H^{(in)} \in \mathbb{R}^{B \times 2 \times (L+1) \times D}$, where $B$ is the batch size (i.e., each example contains $2(L+1)$ tokens).

\subsection{Dual-Level Attention Encoder}
The core of Radio-FM is a Transformer-based encoder that hierarchically models intra-channel temporal dependencies and inter-channel correlations.
As illustrated in Fig.~\ref{fig:architecture}, each encoder layer consists of intra-channel MHSA over the temporal dimension, inter-channel MHSA over the channel dimension, and an FFN, each wrapped in a pre-normalization and residual pathway with LayerScale and DropPath.

\subsubsection{Intra-channel Multi-Head Self-Attention}
We employ Multi-Head Self-Attention (MHSA) independently on each channel to capture long-range temporal dependencies.
To accommodate variable sequence lengths and inject positional information, Rotary Positional Embeddings (RoPE) \cite{su2024roformer} are applied to the query $Q$ and key $K$ vectors.
In the pretraining phase, where a high ratio of patches is masked, it is vital to calculate positional embeddings using the \textit{unmasked} grid positions (i.e., original indices in the full sequence) rather than the compressed indices of the visible patches.
This design keeps relative positional relationships consistent under random masking. The output is computed as:
\begin{equation}
\text{Attn}(Q, K, V) = \text{softmax}\left(\frac{\text{RoPE}(Q) \text{RoPE}(K)^T}{\sqrt{d_k}}\right)V
\end{equation}
where $d_k$ is the head dimension.

\subsubsection{Inter-channel Interaction}
Radio signals are characterized by strong correlations between the In-phase and Quadrature components.
To model these inter-dependencies while preserving channel-specific representations, we introduce an Inter-Channel Multi-Head Self-Attention (MHSA) module.
First, we compress the temporal dimension to obtain a holistic representation for each channel. By applying Global Average Pooling (GAP) on the output of the intra-channel MHSA, we compute the channel token $s_c$:
\begin{equation}
s_c = \text{GAP}(H_c) = \frac{1}{L+1} \sum_{t=0}^{L} H_{c,t}
\end{equation}
The resulting tokens are concatenated to form the channel context matrix $S \in \mathbb{R}^{B \times 2 \times D}$.

The inter-channel attention weights $\alpha \in \mathbb{R}^{B \times 2 \times 2}$ are derived via a self-attention mechanism over the channel dimension:
\begin{equation}
\alpha = \text{softmax}\left( \frac{(S W_q) (S W_k)^T}{\sqrt{d_h}} \right)
\end{equation}
where $W_q, W_k$ are learnable projection matrices.
Subsequently, the feature representation is updated by aggregating context from both channels according to $\alpha$. Formally, for each time step $t$, the refined feature $\hat{H}_{c,t}$ is computed as:
\begin{equation}
\hat{H}_{c,t} = H_{c,t} + \sum_{c' \in \{I,Q\}} \alpha_{c,c'} \cdot (H_{c',t} W_v)
\end{equation}
where $W_v$ is the value projection. This design allows the model to dynamically recalibrate the In-phase features using Quadrature information (and vice-versa) based on their global correlation, preserving distinct channel semantics while enabling phase-aware interaction.

To stabilize training and improve generalization, we employ LayerScale \cite{touvron2021going} combined with Stochastic Depth (DropPath) \cite{huang2016deep}.
Each sub-layer (Intra-channel MHSA, Inter-channel Interaction, FFN) adopts a pre-normalization formulation:
\begin{equation}
Y = X + \text{DropPath}\big(\text{diag}(\gamma) \cdot F(\text{RMSNorm}(X))\big)
\end{equation}
where $\text{RMSNorm}$ denotes root mean square normalization, and $\gamma$ is a learnable diagonal matrix initialized to a small value (e.g., $10^{-5}$).

We further implement a progressive DropPath schedule where the drop probability increases linearly with depth.
For the $i$-th layer ($1 \le i \le L_{enc}$), the rate $p_i$ is:
\begin{equation}
p_i = p_{max} \cdot \frac{i-1}{L_{enc} - 1}
\end{equation}
This strategy preserves information flow in shallow layers while imposing stronger regularization on deeper representations.

\subsection{Pretraining Strategy}
As shown in Fig.~\ref{fig:pipeline} (Stage~I: Pretraining), each unlabeled I/Q sequence is processed by patch embedding, channel-independent (CI) masking, the Radio-FM encoder, and a lightweight decoder for masked reconstruction.
To scale pretraining to heterogeneous datasets with variable sequence lengths, we further employ dataset sampling and dynamic batching under a per-device token budget.

\subsubsection{Channel-independent masking}
To leverage large-scale unlabeled data, we adopt a Masked Autoencoder (MAE) objective adapted for radio signals.
Specifically, we apply \textit{Channel-Independent Random Masking} (CI Mask in Fig.~\ref{fig:pipeline}), where masking patterns are sampled independently for the I and Q channels with a high masking ratio (e.g., 75\%).
This design encourages the encoder to exploit both intra-channel context and inter-channel correlations for reconstruction.
The decoder receives only the visible patches and learnable mask tokens.
The training objective is to minimize the Mean Squared Error (MSE) between the reconstructed patches $\hat{y}$ and the original normalized patches $y$ for the masked locations $M$:
\begin{equation}
\mathcal{L}_{\text{MSE}} = \frac{1}{|M|} \sum_{(c, i) \in M} \|\hat{y}_{c,i} - y_{c,i}\|^2_2
\end{equation}

\subsubsection{Efficient multi-dataset training}
Following the ``Dataset Sample \& Dynamic Batch'' stage in Fig.~\ref{fig:pipeline}, we address severe length heterogeneity via dynamic batching.
Instead of a fixed batch size, we enforce a maximum-token budget per device ($C_{\max}$).
For an input with $L$ patches per channel, the batch size is calculated based on the patch tokens (omitting the $\text{[CLS]}$ token overhead for simplicity) and is set to:
\begin{equation}
B = \left\lfloor \frac{C_{\max}}{2L} \right\rfloor
\end{equation}
This mechanism, combined with proportional dataset sampling, improves GPU utilization and stabilizes convergence across diverse radio signal configurations.

\subsection{Downstream Adaptation}
Following Stage~II in Fig.~\ref{fig:pipeline}, we remove the pretraining decoder and transfer the pretrained backbone to downstream classification tasks by attaching a task-specific prototypical head; the encoder is either frozen (linear probing) or fine-tuned end-to-end.
We form a global representation $h_{cls}$ by averaging the two $\text{[CLS]}$ tokens from the I and Q branches, and compute dot-product prototype logits against $P=[p_1;\ldots;p_K]\in\mathbb{R}^{K\times D}$:
\begin{equation}
\text{prob} = \text{softmax}\left(h_{cls} P^T\right)
\end{equation}
where each $p_k$ is a learnable class prototype.

\section{Experiment\label{sec_3}}

\begin{figure}[htbp]
    \centering
    \includegraphics[width=0.85\linewidth]{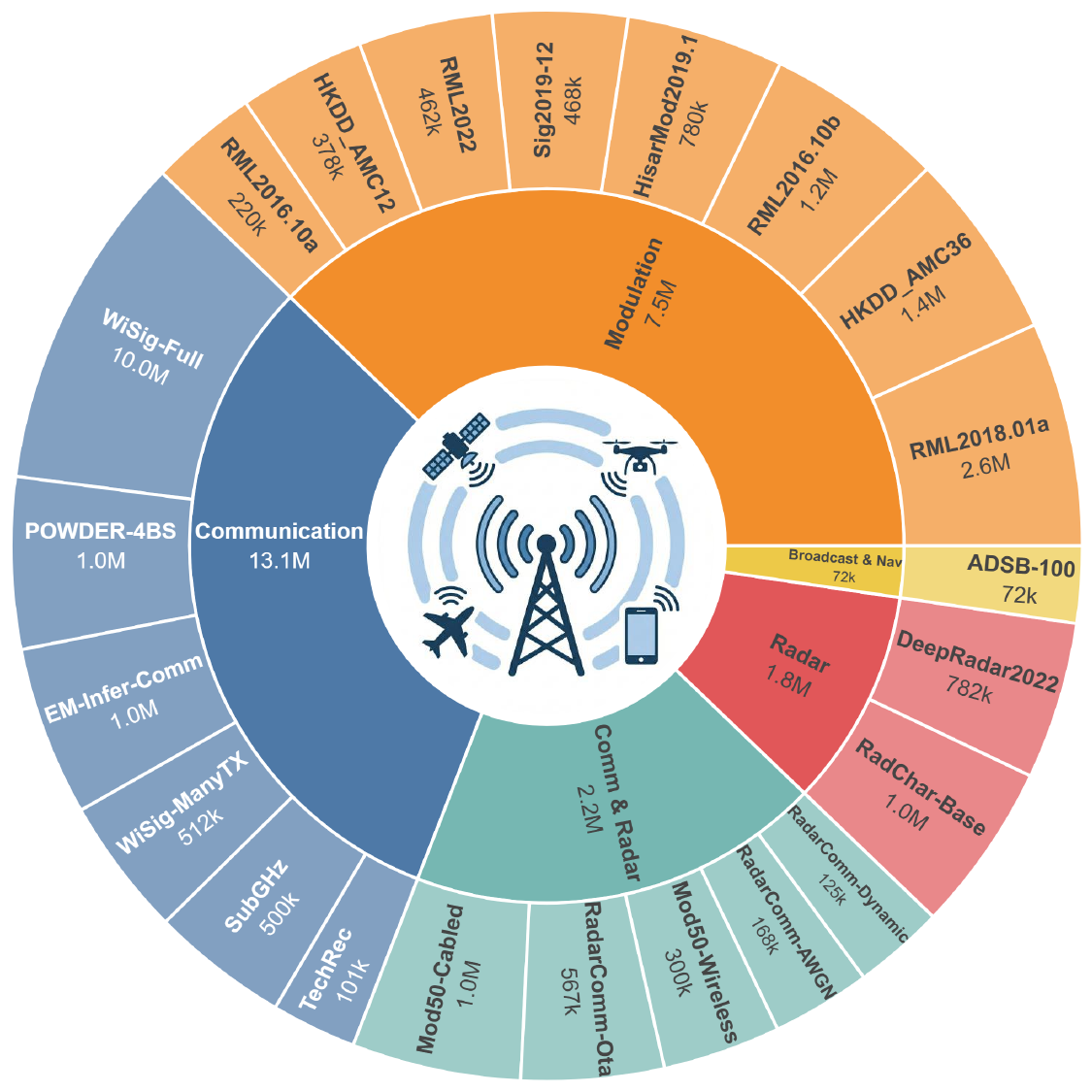}
    \caption{Dataset collection of Radio-FM.}
    \label{fig:dataset_collection}
\end{figure}

\subsection{Setup}

\subsubsection{Datasets}

\begin{table*}[htbp]
    \centering
    \begin{threeparttable}
    \renewcommand{\arraystretch}{1.1} 
    \setlength{\tabcolsep}{6pt} % 稍微减小列间距以容纳新列，防止超宽
    
    \caption{Summary of pretraining datasets}
    \label{tab:pre_training_datasets}
    
    \begin{tabular}{lllcrcr} % 增加了一列，变为 lllllll
    \toprule 
    \textbf{Dataset Name} & \textbf{Signal Type} & \textbf{Task Type} & \textbf{Signal Length} & \textbf{\makecell[l]{Sample Rate / \\ Oversampling}} & \textbf{\makecell[l]{SNR Range \\ (dB)}} & \textbf{Dataset Size} \\ 
    \midrule 
    
    Sig2019-12~\cite{chen2021signet} & Modulation & AMR & 512 & 8$\times$ & -20:2:30 & 468,000 \\
    HKDD\_AMC12~\cite{Zheng10042021} & Modulation & AMR & 512 & 8$\times$ & -20:2:20 & 378,000 \\
    HKDD\_AMC36~\cite{Zheng10042021} & Modulation & AMR & 1024 & 8$\times$ & -20:2:30 & 1,404,000 \\
    HisarMod2019.1~\cite{Tekbiyik2020robust} & Modulation & AMR & 1024 & 2$\times$ & -20:2:18 & 780,000 \\
    RML2018.01a~\cite{rml201801a} & Modulation & AMR & 1024 & 1 MHz & -20:2:30 & 2,555,904 \\
    DeepRadar2022~\cite{10149618} & Radar & RWC & 1024 & 100 MHz & -12:2:20 & 782,000 \\
    WiSig-Full~\cite{9721895} & Communication & SEI & 256 & 25 MHz & N/A & 9,976,477 \\
    POWDER-4BS~\cite{reus2020trust} & Communication & SEI & 512 & 5 MHz & N/A & 1,044,472 \\
    SubGHz~\cite{fontaine2020multi} & Communication & WTR & 2048 & 10 MHz & -15:5:30 & 500,000 \\
    TechRec~\cite{fontaine2019towards} & Communication & WTR & 2048 & 10 MHz & -15:5:30 & 101,380 \\
    Mod50-Cabled\tnote{*} & Radar \& Modulation & AMR \& RWC & 128 & 1 MHz & 6:60 & 1,000,000 \\
    Mod50-Wireless\tnote{*} & Radar \& Modulation & AMR \& RWC & 4096 & 1 MHz & 6:60 & 300,000 \\
    RadarComm-Dynamic~\cite{jagannath2021dataset} & Radar \& Communication & AMR \& RWC \& WTR & 128 & 10 MHz & -20:2:18 & 125,361 \\
    RadarComm-AWGN~\cite{jagannath2021dataset} & Radar \& Communication & AMR \& RWC \& WTR & 128 & 10 MHz & -20:2:18 & 168,000 \\
    RadarComm-Ota2.45GHz~\cite{jagannath2021dataset} & Radar \& Communication & AMR \& RWC \& WTR & 128 & 10 MHz & 0:4:32 & 567,000 \\ 
    
    \bottomrule 
    \end{tabular}
    \begin{tablenotes}
        \footnotesize
        \item[*] Generated using TorchSig~\cite{torchsig}.
    \end{tablenotes}
    \end{threeparttable}
\end{table*}

\begin{table}[htbp]
    \centering
    \begin{threeparttable}
    \renewcommand{\arraystretch}{1.1}

    \caption{Summary of downstream task datasets}
    \label{tab:downstream_datasets}
    
    \begin{tabular}{lccc}
    \toprule 
    \textbf{Dataset Name} & \textbf{Task Type} & \textbf{Signal Length} & \textbf{\makecell[l]{No. of \\ Classes}} \\ 
    \midrule 
    
    RML2016.10a~\cite{o2016convolutional} & AMR & 128 & 11 \\
    RML2016.10b~\cite{o2016convolutional} & AMR & 128 & 10 \\
    RML2022~\cite{Sathyanarayanan2023rml22} & AMR & 128 & 11 \\
    RML2018.01a~\cite{rml201801a} & AMR & 1024 & 24 \\
    Sig2019-12~\cite{chen2021signet} & AMR & 512 & 12 \\
    HKDD\_AMC12~\cite{Zheng10042021} & AMR & 512 & 12 \\
    HKDD\_AMC36~\cite{Zheng10042021} & AMR & 1024 & 36 \\
    HisarMod2019.1~\cite{Tekbiyik2020robust} & AMR & 1024 & 26 \\
    RadChar-Base~\cite{huang2023multi} & RWC & 512 & 5 \\
    DeepRadar2022~\cite{10149618} & RWC & 1024 & 23 \\
    WiSig-ManyTX~\cite{9721895} & SEI & 256 & 150 \\
    ADSB-100\tnote{*}~\cite{ya2022large} & SEI & 4096 & 100 \\
    SubGHz~\cite{fontaine2020multi} & WTR & 2048 & 6 \\
    TechRec~\cite{fontaine2019towards} & WTR & 2048 & 3 \\
    EM-Infer-Comm~\cite{luo2025emind} & WII & 1024 & 9 \\
    
    \bottomrule 
    \end{tabular}
    \begin{tablenotes}
        \footnotesize
        \item[*] Preprocessed by class resampling and ICAO header removal.
    \end{tablenotes}
    \end{threeparttable}
\end{table}

\begin{table*}[htbp]
    \caption{Details of Radio-FM variants. FLOPs are calculated with a sequence length of 1024.}
    \label{tab:model_variants}
    \centering
    \renewcommand{\arraystretch}{1.1}
    \begin{tabular}{lcccccccccc}
    \toprule
    \multirow{2}{*}[-1ex]{Model} & \multicolumn{3}{c}{Encoder Config} & \multicolumn{3}{c}{Decoder Config} & \multicolumn{3}{c}{Parameters} & \multirow{2}{*}[-1ex]{FLOPs} \\
    \cmidrule(lr){2-4} \cmidrule(lr){5-7} \cmidrule(lr){8-10}
    & L & D & H & L & D & H & Encoder & Decoder & Total & \\
    \midrule
    Radio-FM-Tiny & 6 & 128 & 4 & 2 & 64 & 4 & 2.24 M & 0.14 M & 2.38 M & 0.49 G \\
    Radio-FM-Small & 8 & 192 & 6 & 3 & 96 & 6 & 6.32 M & 0.46 M & 6.79 M & 1.48 G \\
    Radio-FM-Base & 12 & 256 & 8 & 4 & 128 & 8 & 16.31 M & 1.08 M & 17.39 M & 3.85 G \\
    Radio-FM-Large & 16 & 384 & 12 & 6 & 192 & 12 & 48.09 M & 3.62 M & 51.71 M & 11.71 G \\
    Radio-FM-XLarge & 20 & 512 & 16 & 8 & 256 & 16 & 106.12 M & 8.53 M & 114.64 M & 26.26 G \\ 
    \bottomrule
    \end{tabular}
\end{table*}

To evaluate Radio-FM's representation learning capabilities, we employ a comprehensive suite of radio signal datasets.
The pretraining phase utilizes 15 diverse datasets spanning modulation, radar, and communication signals (Fig.~\ref{fig:dataset_collection} and Table~\ref{tab:pre_training_datasets}).
For downstream tasks, we evaluate on 15 datasets covering Automatic Modulation Recognition (AMR), Radio Waveform Classification (RWC), Specific Emitter Identification (SEI), and Wireless Technology Recognition (WTR), as detailed in Table~\ref{tab:downstream_datasets}.
Regarding data splitting, we prioritize the official training/test sets provided by the original benchmarks to ensure fair comparison. For datasets without predefined splits, we adopt a default 4:1 stratified random split.

\subsubsection{Model Variants}
We design five Radio-FM variants with varying capacities to investigate scaling behavior.
Inheriting standard Transformer design principles, our models are adapted for 1D radio signals with a fixed patch size of 8. Table~\ref{tab:model_variants} details the architectural specifications, parameter breakdowns, and FLOPs for each variant.

\subsubsection{Pretraining}
We pretrain all variants using a masked reconstruction objective on the combined datasets.
Optimization employs AdamW ($\beta_1=0.9, \beta_2=0.999$, weight decay 0.05) for 10 epochs with a global batch size of 4096.
We schedule the learning rate ($2 \times 10^{-4}$) with cosine decay and a 3\% linear warmup.
Channel-independent masking is applied with a 60\% ratio.

\subsubsection{Evaluation Protocols and Implementation Details}
We employ three protocols to rigorously assess model capabilities across different data regimes:

\begin{itemize}
    \item \textbf{Full Fine-tuning (FT):}
    We update the entire model on the complete training set to benchmark Radio-FM against baselines (SpectrumFM, EMind).
    All models are fine-tuned for 25 epochs using AdamW with a weight decay of 0.01.
    We set the learning rate to $1 \times 10^{-4}$ for Radio-FM (batch size 512) and EMind, while SpectrumFM employs $1 \times 10^{-3}$ following its official configuration.

    \item \textbf{Few-shot Fine-tuning (Few-shot FT):}
    To evaluate data efficiency, we fine-tune the entire model on balanced subsets using a hybrid sampling strategy. 
    Specifically, for datasets stratified by Signal-to-Noise Ratio (SNR), we sample $K \in \{10, 50, 100\}$ instances per SNR level to ensure diverse channel coverage; for datasets without SNR labels, we randomly sample $K$ instances per class.
    This protocol tests the transferability of pretrained representations under varying degrees of label scarcity and channel quality.

    \item \textbf{Linear Probing (LP):}
    Focused on ablation studies (e.g., Mask Ratio, Patch Size), this protocol freezes the encoder to strictly evaluate representation transferability.
    We train a linear head on 100-shot splits for 10 epochs using AdamW (LR $1 \times 10^{-3}$).
\end{itemize}

\subsection{Comparison to State of the Art}
In this subsection, we conduct a comprehensive evaluation of Radio-FM against leading state-of-the-art baselines, including SpectrumFM~\cite{zhou2025spectrumfm} and EMind~\cite{luo2025emind}, across diverse downstream scenarios.

\subsubsection{Full Fine-tuning Comparison}

\begin{table*}[htbp]
    \centering
\caption{Comparison of Full Fine-Tuning Performance on Various Downstream Tasks. The results are reported as \textit{Accuracy}\,/\,\textit{Precision}\,/\,\textit{F1}.}    \label{tab:full_finetuning}
    \renewcommand{\arraystretch}{1.1}
    \begin{tabular}{c|c|cc|ccc}
    \toprule
    \multirow{2}{*}[-1ex]{Task} & \multirow{2}{*}[-1ex]{Dataset} & \multicolumn{2}{c|}{Baselines} & \multicolumn{3}{c}{Radio-FM} \\
    \cmidrule(lr){3-4} \cmidrule(lr){5-7}
    & & SpectrumFM & EMind & Base & Large & XLarge \\
    \midrule
    \multirow{8}{*}{AMR} & RML2016.10a & \textbf{63.75}\,/\,73.44\,/\,65.29 & 60.95\,/\,69.26\,/\,63.01 & \underline{63.50}\,/\,\textbf{74.98}\,/\,\textbf{65.81} & 63.24\,/\,74.53\,/\,65.37 & 63.32\,/\,\underline{74.93}\,/\,\underline{65.68} \\
     & RML2016.10b & 64.33\,/\,\underline{70.01}\,/\,64.44 & 64.96\,/\,68.29\,/\,\textbf{65.77} & \underline{65.23}\,/\,67.14\,/\,65.32 & \textbf{65.27}\,/\,65.46\,/\,65.27 & 65.10\,/\,\textbf{72.32}\,/\,\underline{65.48} \\
     & RML2022 & \textbf{68.16}\,/\,\textbf{71.22}\,/\,\textbf{68.59} & 64.04\,/\,70.10\,/\,64.81 & \underline{66.84}\,/\,70.67\,/\,67.57 & 66.58\,/\,70.20\,/\,67.29 & 67.01\,/\,\underline{71.16}\,/\,\underline{67.76} \\
     & RML2018.01a & 56.57\,/\,60.78\,/\,57.18 & 63.55\,/\,65.46\,/\,63.64 & \underline{64.32}\,/\,\textbf{73.41}\,/\,\underline{66.24} & 64.22\,/\,71.54\,/\,65.92 & \textbf{64.70}\,/\,\underline{72.49}\,/\,\textbf{66.53} \\
     & Sig2019-12 & 62.08\,/\,62.83\,/\,62.04 & 66.23\,/\,67.18\,/\,66.47 & \textbf{71.10}\,/\,72.35\,/\,71.07 & \underline{70.93}\,/\,\textbf{74.86}\,/\,\textbf{71.50} & 70.81\,/\,\underline{74.26}\,/\,\underline{71.44} \\
     & HKDD\_AMC12 & 54.46\,/\,55.24\,/\,54.45 & 51.92\,/\,52.99\,/\,52.08 & 63.06\,/\,64.08\,/\,63.08 & \textbf{63.08}\,/\,\textbf{66.07}\,/\,\textbf{63.47} & \underline{62.92}\,/\,\underline{64.03}\,/\,\underline{62.96} \\
     & HKDD\_AMC36 & 55.52\,/\,57.73\,/\,55.34 & 58.06\,/\,58.40\,/\,57.55 & \textbf{67.33}\,/\,\textbf{68.42}\,/\,\textbf{67.16} & 65.75\,/\,67.85\,/\,\underline{65.36} & \underline{65.92}\,/\,\underline{68.09}\,/\,64.84 \\
     & HisarMod2019.1 & 59.42\,/\,59.25\,/\,58.87 & \underline{79.32}\,/\,79.32\,/\,\underline{79.30} & 78.30\,/\,78.45\,/\,78.32 & 80.53\,/\,\underline{80.67}\,/\,80.56 & \textbf{80.90}\,/\,\textbf{81.01}\,/\,\textbf{80.92} \\
    \midrule
    \multirow{2}{*}{RWC} & RadChar-Base & 88.37\,/\,88.34\,/\,88.35 & 89.65\,/\,89.64\,/\,89.64 & \textbf{90.13}\,/\,\textbf{90.13}\,/\,\textbf{90.12} & 90.06\,/\,90.06\,/\,90.04 & \underline{90.12}\,/\,\underline{90.13}\,/\,\underline{90.12} \\
     & DeepRadar2022 & 69.63\,/\,69.93\,/\,69.34 & 79.76\,/\,80.08\,/\,79.85 & \underline{87.07}\,/\,\underline{87.18}\,/\,\underline{87.02} & 86.61\,/\,87.01\,/\,86.64 & \textbf{87.45}\,/\,\textbf{87.58}\,/\,\textbf{87.48} \\
    \midrule
    \multirow{2}{*}{SEI} & WiSig-ManyTX & 85.01\,/\,87.40\,/\,86.17 & 91.48\,/\,91.54\,/\,90.80 & 91.37\,/\,\textbf{93.43}\,/\,\textbf{92.16} & \textbf{91.64}\,/\,\underline{92.95}\,/\,\underline{92.14} & \underline{91.61}\,/\,92.28\,/\,91.87 \\
     & ADSB-100 & 26.89\,/\,25.68\,/\,24.81 & \underline{99.26}\,/\,\underline{99.27}\,/\,\underline{99.25} & 96.85\,/\,96.91\,/\,96.85 & 99.19\,/\,99.21\,/\,99.19 & \textbf{99.41}\,/\,\textbf{99.42}\,/\,\textbf{99.41} \\
    \midrule
    \multirow{2}{*}{WTR} & SubGHz & 78.76\,/\,70.16\,/\,76.33 & 83.32\,/\,82.19\,/\,58.93 & \underline{84.23}\,/\,\underline{83.34}\,/\,\underline{81.89} & \textbf{84.60}\,/\,\textbf{83.84}\,/\,\textbf{82.34} & 84.41\,/\,83.58\,/\,82.10 \\
     & TechRec & 76.99\,/\,77.34\,/\,77.05 & 86.45\,/\,86.63\,/\,86.34 & \textbf{89.70}\,/\,\underline{90.20}\,/\,\textbf{89.78} & \underline{89.64}\,/\,\textbf{90.35}\,/\,\underline{89.73} & 89.44\,/\,89.74\,/\,89.50 \\
    \midrule
    WII & EM-Infer-Comm & 65.25\,/\,68.36\,/\,65.72 & 82.40\,/\,83.15\,/\,82.53 & \textbf{84.02}\,/\,\textbf{84.31}\,/\,\textbf{83.93} & \underline{83.93}\,/\,\textbf{84.20}\,/\,\underline{83.93} & 83.72\,/\,83.96\,/\,83.74 \\
    \bottomrule
    \end{tabular}
\end{table*}

Table~\ref{tab:full_finetuning} details the full fine-tuning performance, highlighting Radio-FM's adaptability across diverse signal complexities.
Radio-FM achieves state-of-the-art results on 13 of 15 datasets, with particular dominance in tasks requiring fine-grained physical interpretation (SEI, WTR, WII).
We attribute this success to two architectural advantages. First, unlike SpectrumFM, which is constrained by fixed-length inputs (truncating signals $>128$), Radio-FM captures long-range temporal dependencies, explaining its massive lead on long-sequence datasets like SubGHz.
Second, the dual-channel processing specifically optimized for I/Q independence models signal dependencies better than standard transformers (EMind), enabling precise characterization of complex waveforms---evidenced by the +7.69\% and +9.27\% gains on DeepRadar2022 and HKDD\_AMC36, respectively.
Furthermore, we observe a task-dependent scaling law: while performance saturates on simpler, low-entropy tasks (e.g., RML2016), capacity scaling remains critical for high-fidelity benchmarks.
For instance, moving from Base to XLarge yields continuous gains on ADSB-100 and HisarMod2019.1, confirming that larger models are essential for modeling the high variability of real-world radio environments.

\begin{figure*}[htbp]
    \centering
    \captionsetup[subfigure]{skip=1pt}
    % --- Row 1: prerained ---
    \begin{subfigure}{0.19\textwidth}
        \centering
        \includegraphics[width=\linewidth]{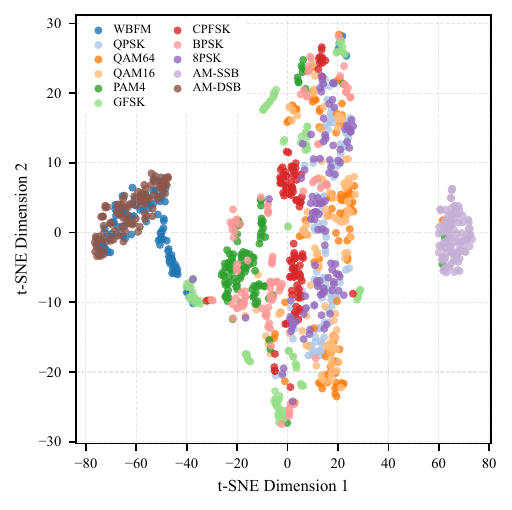}
        \caption{SpectrumFM (Pre)}
        \label{fig:tsne_spec_pre}
    \end{subfigure}
    \hfill
    \begin{subfigure}{0.19\textwidth}
        \centering
        \includegraphics[width=\linewidth]{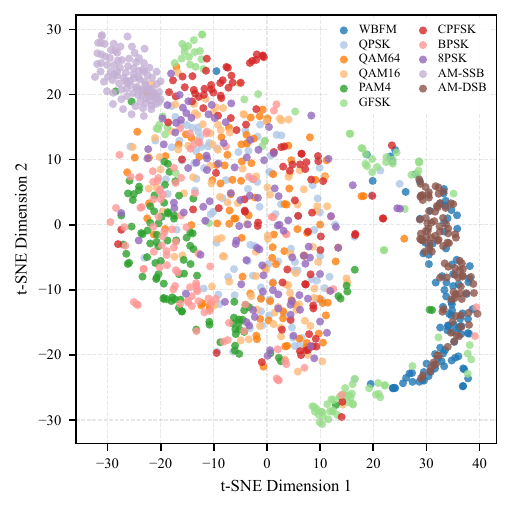}
        \caption{EMind (Pre)}
        \label{fig:tsne_emind_pre}
    \end{subfigure}
    \hfill
    \begin{subfigure}{0.19\textwidth}
        \centering
        \includegraphics[width=\linewidth]{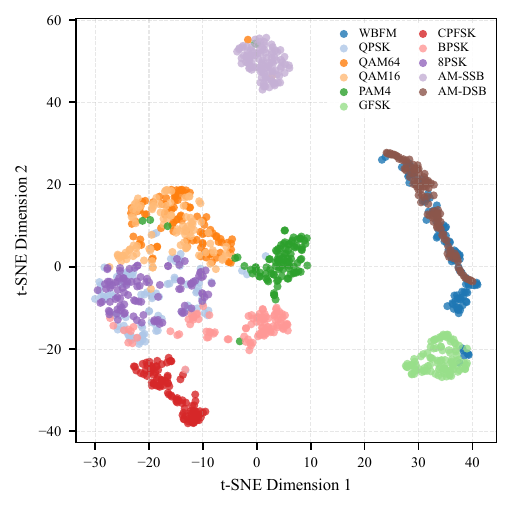}
        \caption{Radio-FM-B (Pre)}
        \label{fig:tsne_radiofm_base_pre}
    \end{subfigure}
    \hfill
    \begin{subfigure}{0.19\textwidth}
        \centering
        \includegraphics[width=\linewidth]{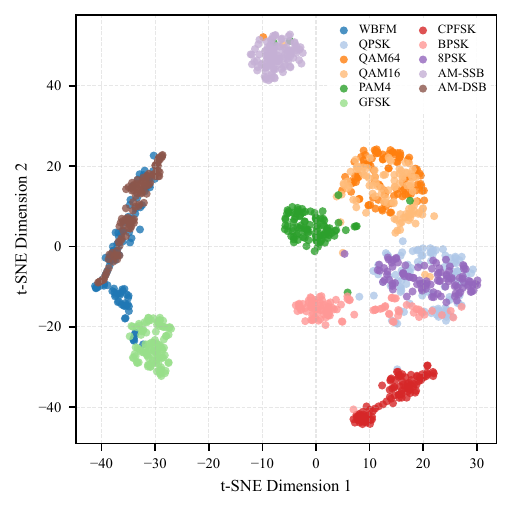}
        \caption{Radio-FM-L (Pre)}
        \label{fig:tsne_radiofm_large_pre}
    \end{subfigure}
    \hfill
    \begin{subfigure}{0.19\textwidth}
        \centering
        \includegraphics[width=\linewidth]{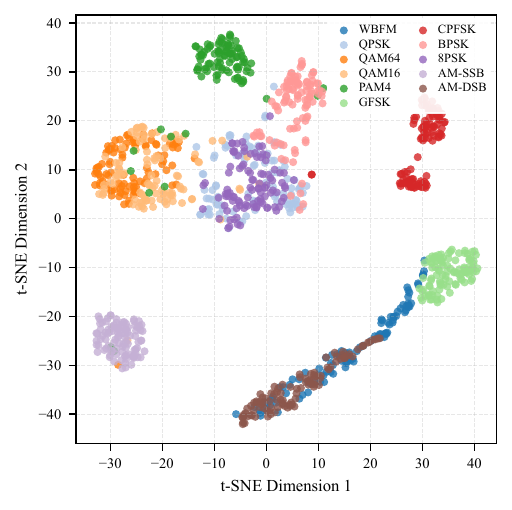}
        \caption{Radio-FM-XL (Pre)}
        \label{fig:tsne_radiofm_xlarge_pre}
    \end{subfigure}
    
    % \vspace{0.3cm} % vertical spacing
    
    % --- Row 2: Fine-tuned ---
    \begin{subfigure}{0.19\textwidth}
        \centering
        \includegraphics[width=\linewidth]{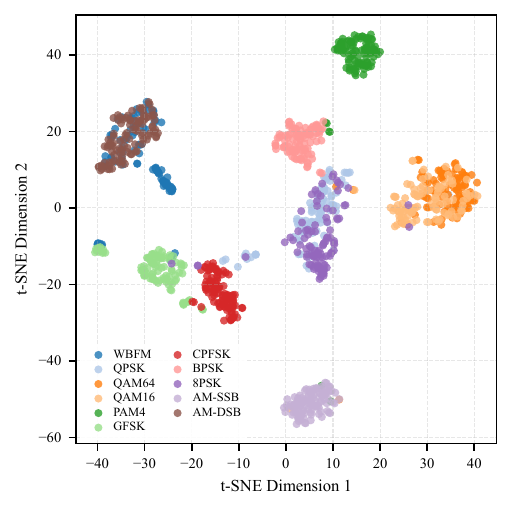}
        \caption{SpectrumFM (FT)}
        \label{fig:tsne_spec_ft}
    \end{subfigure}
    \hfill
    \begin{subfigure}{0.19\textwidth}
        \centering
        \includegraphics[width=\linewidth]{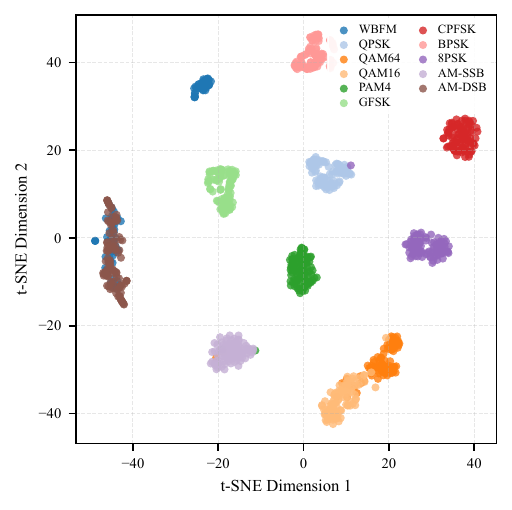}
        \caption{EMind (FT)}
        \label{fig:tsne_emind_ft}
    \end{subfigure}
    \hfill
    \begin{subfigure}{0.19\textwidth}
        \centering
        \includegraphics[width=\linewidth]{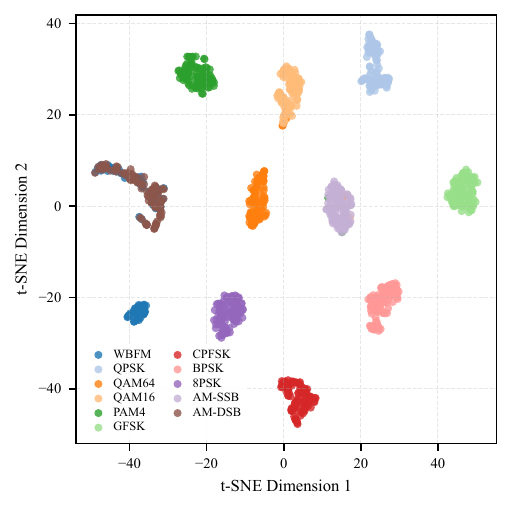}
        \caption{Radio-FM-B (FT)}
        \label{fig:tsne_radiofm_base_ft}
    \end{subfigure}
    \hfill
    \begin{subfigure}{0.19\textwidth}
        \centering
        \includegraphics[width=\linewidth]{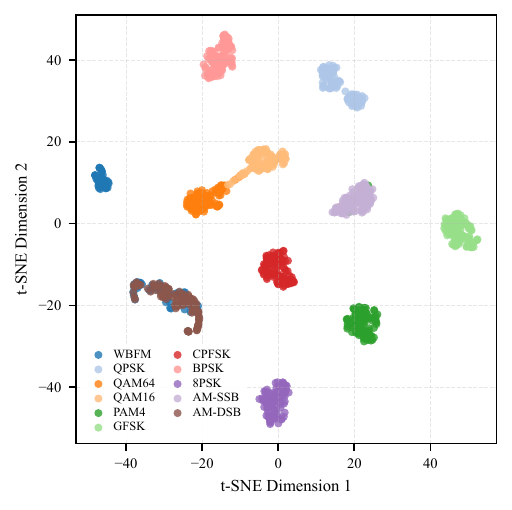}
        \caption{Radio-FM-L (FT)}
        \label{fig:tsne_radiofm_large_ft}
    \end{subfigure}
    \hfill
    \begin{subfigure}{0.19\textwidth}
        \centering
        \includegraphics[width=\linewidth]{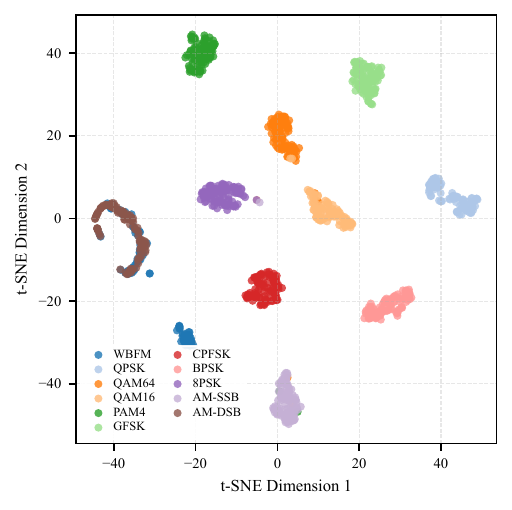}
        \caption{Radio-FM-XL (FT)}
        \label{fig:tsne_radiofm_xlarge_ft}
    \end{subfigure}
    
    \caption{t-SNE visualization on RML2016.10a (AMR task) at 10dB SNR. From left to right: SpectrumFM, EMind, Radio-FM Base, Large, and XLarge. Top row: Pretrained; Bottom row: Full Fine-tuned.}
    \label{fig:tsne_comparison}
\end{figure*}

\begin{figure*}[htbp]
    \centering
    \captionsetup[subfigure]{skip=1pt}
    % --- Row 1: Pretrained ---
    \begin{subfigure}{0.19\textwidth}
        \centering
        \includegraphics[width=\linewidth]{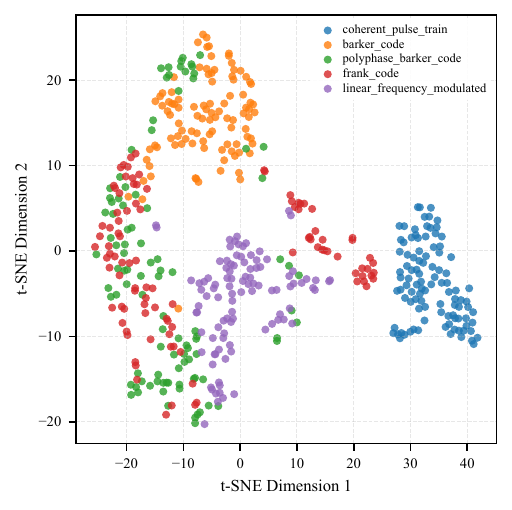}
        \caption{SpectrumFM (Pre)}
        \label{fig:tsne_radchar_spec_pre}
    \end{subfigure}
    \hfill
    \begin{subfigure}{0.19\textwidth}
        \centering
        \includegraphics[width=\linewidth]{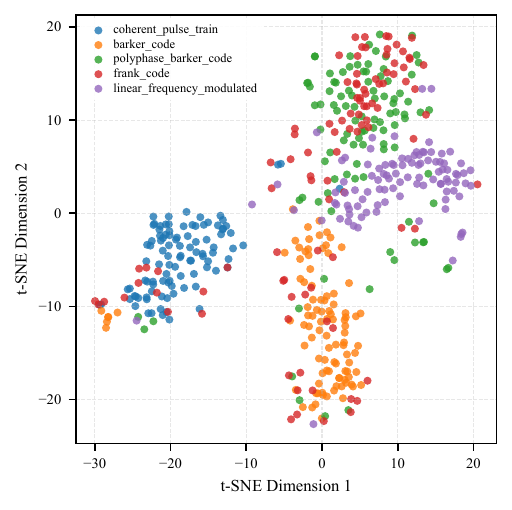}
        \caption{EMind (Pre)}
        \label{fig:tsne_radchar_emind_pre}
    \end{subfigure}
    \hfill
    \begin{subfigure}{0.19\textwidth}
        \centering
        \includegraphics[width=\linewidth]{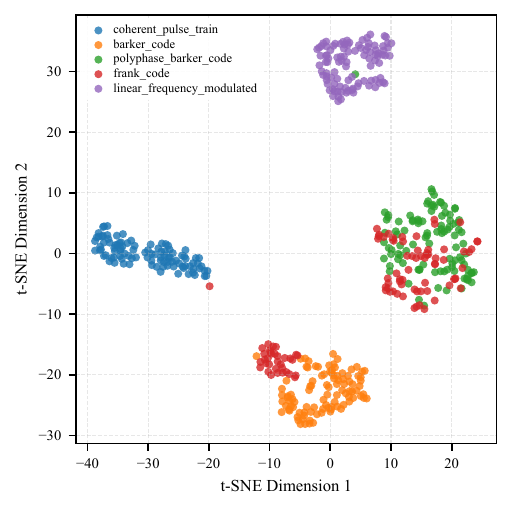}
        \caption{Radio-FM-B (Pre)}
        \label{fig:tsne_radchar_radiofm_base_pre}
    \end{subfigure}
    \hfill
    \begin{subfigure}{0.19\textwidth}
        \centering
        \includegraphics[width=\linewidth]{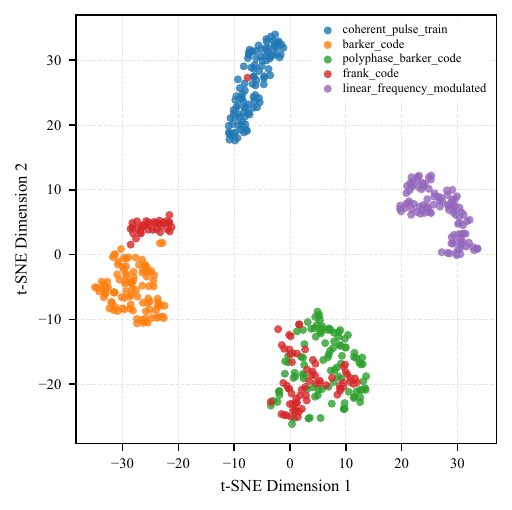}
        \caption{Radio-FM-L (Pre)}
        \label{fig:tsne_radchar_radiofm_large_pre}
    \end{subfigure}
    \hfill
    \begin{subfigure}{0.19\textwidth}
        \centering
        \includegraphics[width=\linewidth]{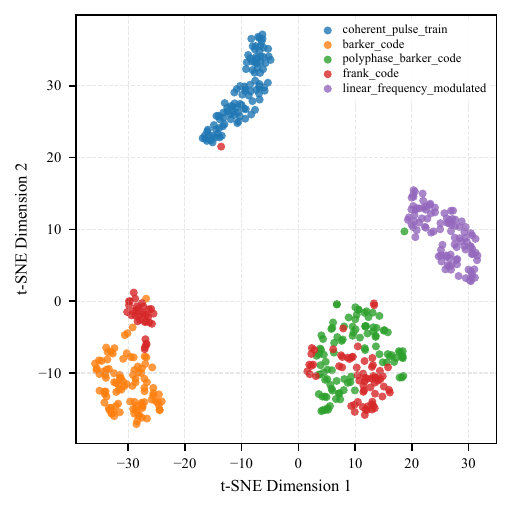}
        \caption{Radio-FM-XL (Pre)}
        \label{fig:tsne_radchar_radiofm_xlarge_pre}
    \end{subfigure}
    
    % \vspace{0.3cm}
    
    % --- Row 2: Fine-tuned ---
    \begin{subfigure}{0.19\textwidth}
        \centering
        \includegraphics[width=\linewidth]{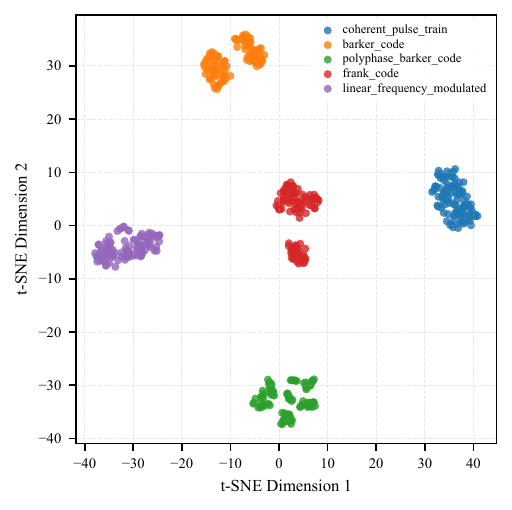}
        \caption{SpectrumFM (FT)}
        \label{fig:tsne_radchar_spec_ft}
    \end{subfigure}
    \hfill
    \begin{subfigure}{0.19\textwidth}
        \centering
        \includegraphics[width=\linewidth]{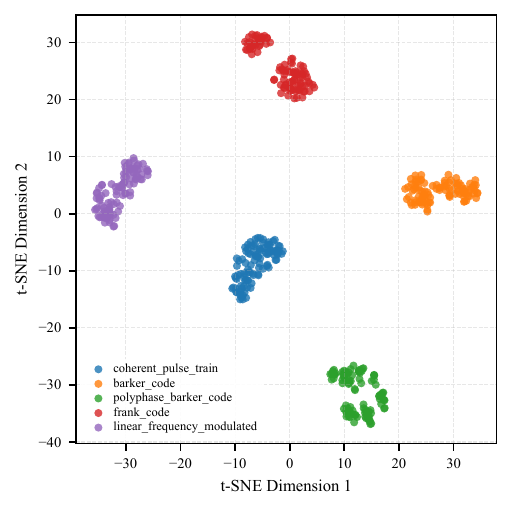}
        \caption{EMind (FT)}
        \label{fig:tsne_radchar_emind_ft}
    \end{subfigure}
    \hfill  
    \begin{subfigure}{0.19\textwidth}
        \centering
        \includegraphics[width=\linewidth]{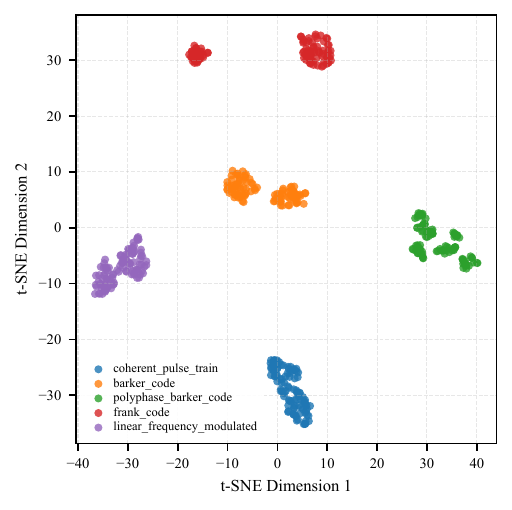}
        \caption{Radio-FM-B (FT)}
        \label{fig:tsne_radchar_radiofm_base_ft}
    \end{subfigure}
    \hfill  
    \begin{subfigure}{0.19\textwidth}
        \centering
        \includegraphics[width=\linewidth]{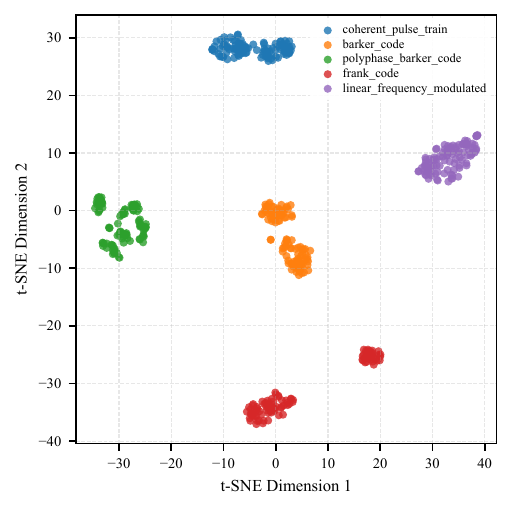}
        \caption{Radio-FM-L (FT)}
        \label{fig:tsne_radchar_radiofm_large_ft}
    \end{subfigure}
    \hfill  
    \begin{subfigure}{0.19\textwidth}
        \centering
        \includegraphics[width=\linewidth]{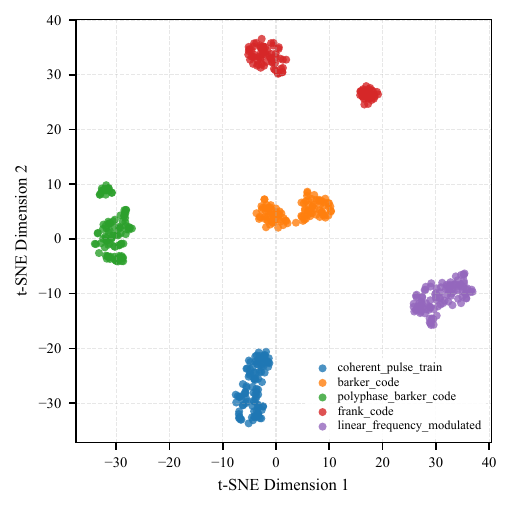}
        \caption{Radio-FM-XL (FT)}
        \label{fig:tsne_radchar_radiofm_xlarge_ft}
    \end{subfigure}
    
    \caption{t-SNE visualization on RadChar-Base (RWC task) at 10dB SNR. From left to right: SpectrumFM, EMind, Radio-FM Base, Large, and XLarge. Top row: Pretrained; Bottom row: Full Fine-tuned.}
    \label{fig:tsne_radchar_comparison}
\end{figure*}

\begin{figure*}[htbp]
    \centering
    \captionsetup[subfigure]{skip=1pt}
    % --- Row 1: Pretrained ---
    \begin{subfigure}{0.19\textwidth}
        \centering
        \includegraphics[width=\linewidth]{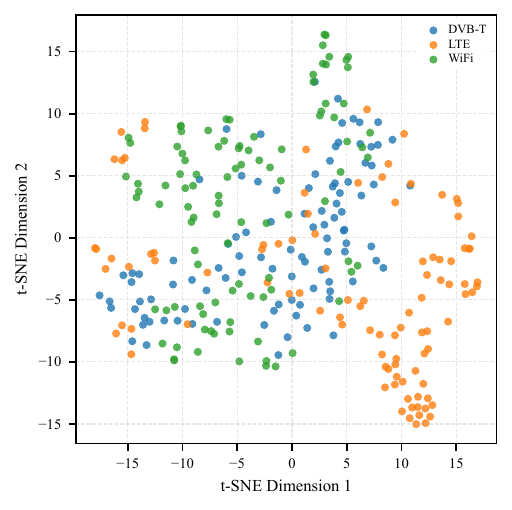}
        \caption{SpectrumFM (Pre)}
        \label{fig:tsne_techrec_spec_pre}
    \end{subfigure}
    \hfill
    \begin{subfigure}{0.19\textwidth}
        \centering
        \includegraphics[width=\linewidth]{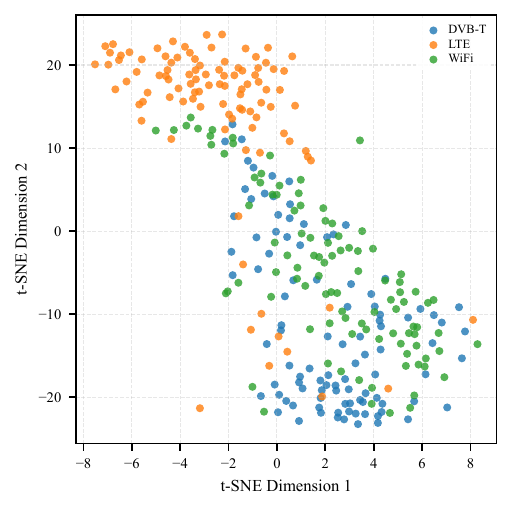}
        \caption{EMind (Pre)}
        \label{fig:tsne_techrec_emind_pre}
    \end{subfigure}
    \hfill
    \begin{subfigure}{0.19\textwidth}
        \centering
        \includegraphics[width=\linewidth]{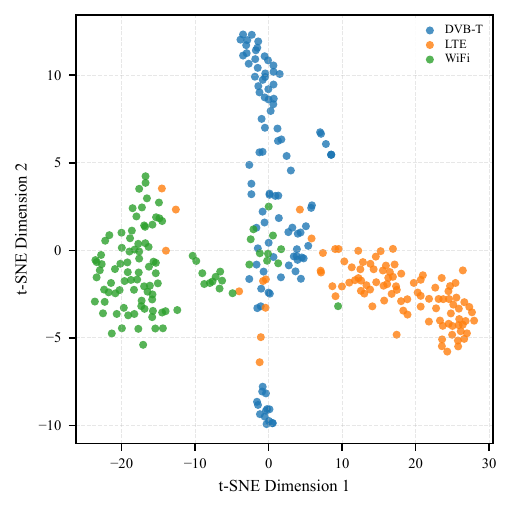}
        \caption{Radio-FM-B (Pre)}
        \label{fig:tsne_techrec_radiofm_base_pre}
    \end{subfigure}
    \hfill
    \begin{subfigure}{0.19\textwidth}
        \centering
        \includegraphics[width=\linewidth]{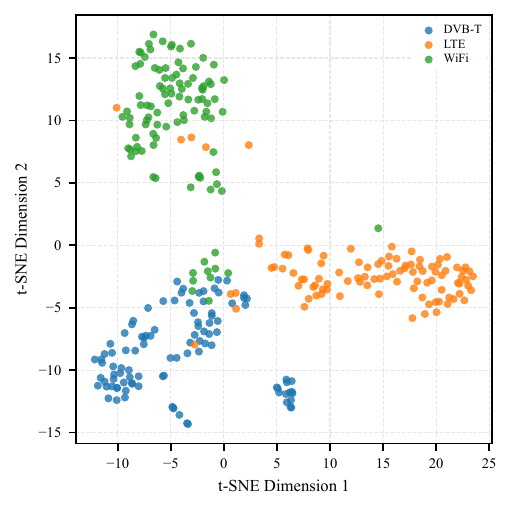}
        \caption{Radio-FM-L (Pre)}
        \label{fig:tsne_techrec_radiofm_large_pre}
    \end{subfigure}
    \hfill
    \begin{subfigure}{0.19\textwidth}
        \centering
        \includegraphics[width=\linewidth]{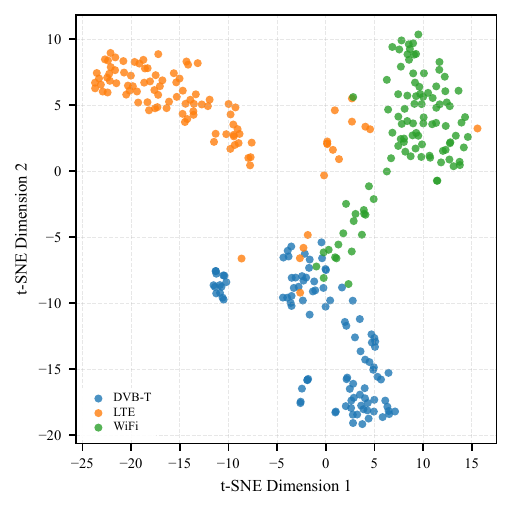}
        \caption{Radio-FM-XL (Pre)}
        \label{fig:tsne_techrec_radiofm_xlarge_pre}
    \end{subfigure}
    
    % \vspace{0.3cm}
    
    % --- Row 2: Fine-tuned ---
    \begin{subfigure}{0.19\textwidth}
        \centering
        \includegraphics[width=\linewidth]{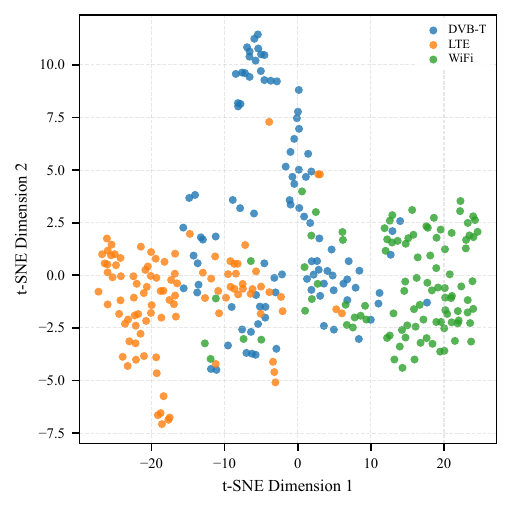}
        \caption{SpectrumFM (FT)}
        \label{fig:tsne_techrec_spec_ft}
    \end{subfigure}
    \hfill
    \begin{subfigure}{0.19\textwidth}
        \centering
        \includegraphics[width=\linewidth]{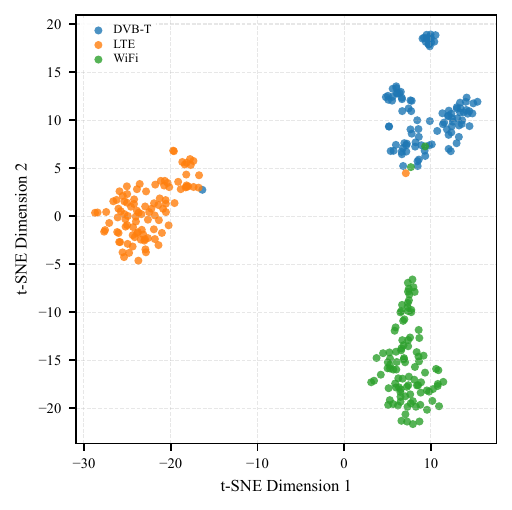}
        \caption{EMind (FT)}
        \label{fig:tsne_techrec_emind_ft}
    \end{subfigure}
    \hfill
    \begin{subfigure}{0.19\textwidth}
        \centering
        \includegraphics[width=\linewidth]{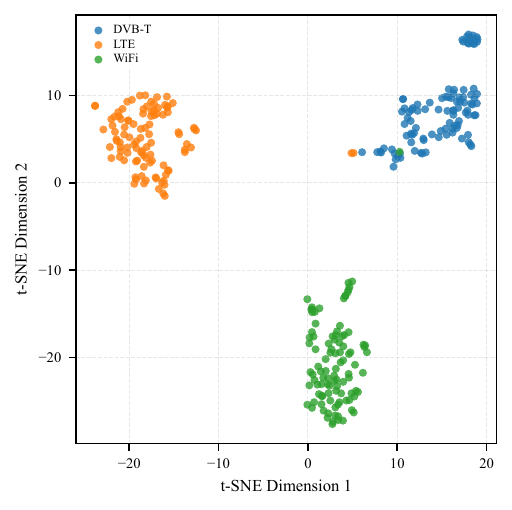}
        \caption{Radio-FM-B (FT)}
        \label{fig:tsne_techrec_radiofm_base_ft}
    \end{subfigure}
    \hfill
    \begin{subfigure}{0.19\textwidth}
        \centering
        \includegraphics[width=\linewidth]{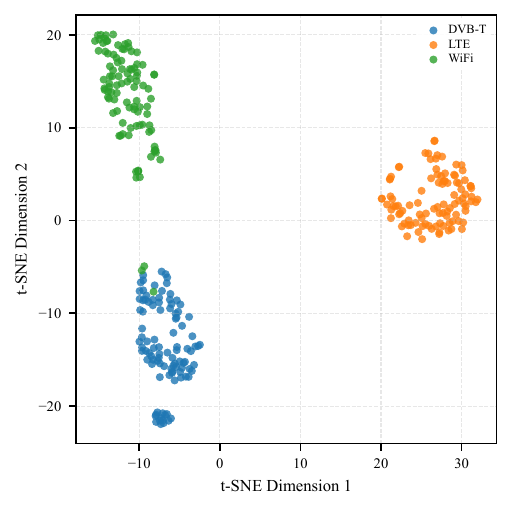}
        \caption{Radio-FM-L (FT)}
        \label{fig:tsne_techrec_radiofm_large_ft}
    \end{subfigure}
    \hfill
    \begin{subfigure}{0.19\textwidth}
        \centering
        \includegraphics[width=\linewidth]{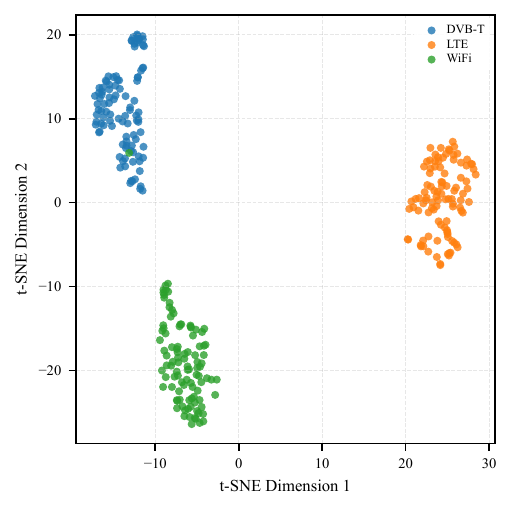}
        \caption{Radio-FM-XL (FT)}
        \label{fig:tsne_techrec_radiofm_xlarge_ft}
    \end{subfigure}
    
    \caption{t-SNE visualization on TechRec (WTR task) at 10dB SNR. From left to right: SpectrumFM, EMind, Radio-FM Base, Large, and XLarge. Top row: Pretrained; Bottom row: Full Fine-tuned.}
    \label{fig:tsne_techrec_comparison}
\end{figure*}

To qualitatively validate representation quality, we visualize feature manifolds across three representative domains: AMR (Fig.~\ref{fig:tsne_comparison}), RWC (Fig.~\ref{fig:tsne_radchar_comparison}), and WTR (Fig.~\ref{fig:tsne_techrec_comparison}).
We specifically select these three datasets from the 15 benchmarks as they exemplify canonical electromagnetic pattern recognition tasks and have been consistently evaluated by the baselines.
To ensure a fair assessment of pretraining alignment (top rows), we extract features directly from the encoder output for both Radio-FM and EMind, avoiding any bias from task-specific heads.
The visualizations reveal that while baseline pretrained features are largely entangled (e.g., Fig.~\ref{fig:tsne_comparison}a-b), Radio-FM exhibits emergent structural organization even without supervision (Fig.~\ref{fig:tsne_comparison}c-e).
This superior initialization directly amplifies fine-tuning effectiveness.
As seen in the bottom rows, Radio-FM produces significantly tighter and more distinct clusters.
For instance, in the complex RML2016.10a task, baselines fail to separate high-order modulations (e.g., QAM16 vs. QAM64), whereas Radio-FM successfully disentangles these classes.
This enhanced separability confirms that our dual-channel processing pretraining strategy facilitates the learning of discriminative, low-entropy representations, underpinning the SOTA performance in Table~\ref{tab:full_finetuning}.

\subsubsection{Few-shot Learning Comparison}

\begin{table*}[h]
    \centering
    \caption{Comparison of few-shot fine-tuning accuracy (\%) on Various Downstream Tasks.}	    
    \label{tab:few_shot}
    \renewcommand{\arraystretch}{1.1}
    \setlength{\tabcolsep}{1.5pt}
    \begin{tabular}{c|c|ccccc|ccccc|ccccc}
    \toprule
    \multirow{2}{*}[-1ex]{Task} & \multirow{2}{*}[-1ex]{Dataset} & \multicolumn{5}{c|}{10-Shot} & \multicolumn{5}{c|}{50-Shot} & \multicolumn{5}{c}{100-Shot} \\
    \cmidrule(lr){3-7} \cmidrule(lr){8-12} \cmidrule(lr){13-17}
     & & SpecFM & EMind & Ours-B & Ours-L & Ours-XL & SpecFM & EMind & Ours-B & Ours-L & Ours-XL & SpecFM & EMind & Ours-B & Ours-L & Ours-XL \\
    \midrule
    \multirow{8}{*}{AMR} 
    & RML2016.10a     & \textbf{51.01} & 42.52 & 47.20 & 46.33 & 47.22 & \textbf{55.14} & 47.78 & 52.87 & 53.49 & 53.53 & \textbf{61.23} & 51.03 & 55.54 & 56.33 & 56.53 \\
    & RML2016.10b     & \textbf{51.28} & 44.39 & 48.07 & 49.76 & 49.66 & 55.85 & 50.50 & 54.10 & 55.27 & \textbf{56.18} & \textbf{62.18} & 53.41 & 56.60 & 58.05 & 58.00 \\
    & RML2022         & \textbf{52.19} & 41.56 & 48.47 & 50.33 & 50.76 & \textbf{61.15} & 48.40 & 55.01 & 55.28 & 54.50 & \textbf{62.67} & 52.47 & 57.73 & 58.01 & 58.35 \\
    & RML2018.01a     & 35.01 & 34.87 & 40.66 & 43.89 & \textbf{46.70} & 47.32 & 46.32 & 54.20 & 55.51 & \textbf{56.58} & 49.87 & 51.52 & 58.80 & 59.44 & \textbf{59.66} \\
    & Sig2019-12      & 39.71 & 44.11 & 47.26 & 48.46 & \textbf{51.85} & 53.53 & 52.88 & 57.70 & \textbf{59.98} & 59.50 & 56.08 & 55.02 & 60.91 & 62.55 & \textbf{62.92} \\
    & HKDD\_AMC12     & 24.05 & 29.17 & 38.26 & 40.17 & \textbf{40.76} & 40.78 & 42.70 & 46.74 & 47.95 & \textbf{48.25} & 46.35 & 45.02 & 49.49 & 50.60 & \textbf{51.81} \\
    & HKDD\_AMC36     & 36.55 & 38.75 & 46.26 & 46.74 & \textbf{47.75} & 48.03 & 48.43 & 52.09 & 54.56 & \textbf{54.64} & 50.67 & 51.38 & 56.27 & 56.62 & \textbf{57.02} \\
    & HisarMod2019.1  & 33.82 & 35.40 & 41.71 & 43.03 & \textbf{43.13} & 39.88 & 46.56 & 51.04 & 52.30 & \textbf{52.40} & 43.52 & 52.94 & 56.36 & 58.13 & \textbf{58.28} \\
    \midrule
    \multirow{2}{*}{RWC} 
    & RadChar-Base    & \textbf{81.58} & 80.44 & 78.61 & 79.21 & 78.76 & 83.47 & 83.19 & 83.03 & 83.37 & \textbf{83.52} & 84.78 & 84.53 & 84.95 & \textbf{85.02} & 84.88 \\
    & DeepRadar2022   & 28.41 & 40.55 & 49.31 & 51.10 & \textbf{52.46} & 51.14 & 59.36 & 65.89 & 66.95 & \textbf{67.58} & 56.42 & 65.64 & 70.56 & 71.23 & \textbf{71.55} \\
    \midrule
    \multirow{2}{*}{SEI} 
    & WiSig-ManyTX    & 2.58 & 4.63 & 5.48 & 9.39 & \textbf{11.82} & 10.29 & 17.38 & 33.87 & 37.88 & \textbf{38.95} & 20.51 & 32.43 & 46.85 & 50.23 & \textbf{51.99} \\
    & ADSB-100        & 3.04 & \textbf{35.36} & 21.47 & 25.39 & 24.13 & 11.29 & \textbf{76.29} & 75.29 & 76.23 & 75.87 & 17.01 & 89.88 & 90.61 & 91.24 & \textbf{92.33} \\
    \midrule
    \multirow{2}{*}{WTR} 
    & SubGHz          & 51.56 & 52.03 & 77.16 & \textbf{77.36} & 76.33 & 63.11 & 67.55 & 67.26 & \textbf{77.86} & 77.27 & 62.50 & \textbf{79.46} & 76.89 & 76.93 & 70.23 \\
    & TechRec         & 70.77 & 72.55 & 80.94 & 78.32 & \textbf{81.10} & 72.35 & 75.85 & 80.57 & 81.23 & \textbf{83.13} & 72.93 & 77.70 & \textbf{82.58} & 82.53 & 82.27 \\
    \midrule
    WII 
    & EM-Infer-Comm   & 50.58 & 60.37 & 62.35 & 61.72 & \textbf{63.11} & 57.49 & 69.01 & 72.96 & 72.29 & \textbf{73.37} & 59.34 & 72.93 & 75.29 & \textbf{75.71} & 75.46 \\
    \midrule
    \textbf{Best Count} & \textbf{All Tasks} & 
    4 & 1 & 0 & 1 & \textbf{9} & 
    2 & 1 & 0 & 2 & \textbf{10} &
    3 & 1 & 1 & 2 & \textbf{8} \\
    \bottomrule
    \end{tabular}
\end{table*}

Table~\ref{tab:few_shot} underscores the sample efficiency of Radio-FM.
The results reveal a clear dichotomy: while SpectrumFM leverages specialized inductive biases to perform well on standard modulation sets (e.g., the RML series) at 10-shot, it suffers significant performance drops on tasks requiring semantic abstraction beyond simple modulation types.
In contrast, Radio-FM demonstrates robust few-shot adaptation, with the XLarge variant dominating 9 of the 15 datasets in the 10-shot regime.
Most notably, on data-hungry tasks like DeepRadar2022, Radio-FM-XLarge surpasses SpectrumFM by over 24\%, suggesting that its pretraining prior is rich enough to characterize complex radar waveforms from minimal examples.
This advantage scales efficiently: at 100-shot, Radio-FM not only consolidates its lead on challenging benchmarks (e.g., +8.1\% vs. EMind on RML2018) but also enables rapid convergence on emitter identification (ADSB-100), achieving 92.33\% accuracy where baselines struggle to detangle subtle hardware fingerprints.

%     \midrule
%     \multirow{2}{*}{WTR} 
%      & SubGHz & 66.33 & 68.80 & 75.82 & 74.34 & \textbf{76.23} & 67.88 & 66.20 & 63.68 & 71.90 & \textbf{72.34} \\
%      & TechRec & 80.67 & 77.80 & \textbf{81.53} & 78.53 & 76.13 & 77.92 & \textbf{79.03} & 75.17 & 73.90 & 72.79 \\
%     \midrule
%     WII 
%      & EM-Infer-Comm & 49.02 & 48.14 & \textbf{50.55} & 50.09 & 49.03 & 45.42 & 45.96 & 47.62 & \textbf{49.01} & 45.88 \\
%     \midrule
%     Avg & All Tasks & 43.60 & 43.07 & \textbf{44.43} & 43.21 & 40.32 & \textbf{42.44} & 41.90 & 42.32 & 42.40 & 39.22 \\
%     \bottomrule
%     \end{tabular}
% \end{table*}

\subsection{Comprehensive Model Analysis and Ablations}
This section provides an in-depth analysis of Radio-FM, including scaling behavior, hyperparameter sensitivity, and module contributions.
Unless otherwise specified, the metric reported is the arithmetic mean accuracy averaged across all 15 downstream datasets.
Note that while we evaluate scaling laws across the full model family, all subsequent hyperparameter and component ablations (Mask Ratio, Patch Size, and Module Analysis) are conducted using the Radio-FM-Base variant to balance representational capacity with computational efficiency.

\subsubsection{Model Scaling Analysis}

\begin{table*}[htbp]
    \centering
\caption{Consolidated Accuracy Comparison Across Different Training Paradigms. Each cell displays results formatted as: \textit{From Scratch / Few-shot FT (100-shot) / Full Finetune}. The best result among different model scales for each specific setting is marked in \textbf{bold}.}	    \label{tab:consolidated_results}
    \renewcommand{\arraystretch}{1.1}
    \setlength{\tabcolsep}{3pt}
    \begin{tabular}{c|c|ccccc}
    \toprule
    Task Domain & Dataset & RadioFM-Tiny & RadioFM-Small & RadioFM-Base & RadioFM-Large & RadioFM-XLarge \\
    \midrule
    \multirow{8}{*}{AMR} 
     & RML2016.10a & 56.19 / 55.19 / 62.93 & 58.96 / 54.16 / 62.94 & 59.84 / 55.54 / \textbf{63.50} & 60.58 / 56.33 / 63.24 & \textbf{61.35} / \textbf{56.53} / 63.32 \\
     & RML2016.10b & 64.83 / 55.32 / \textbf{65.32} & 65.40 / 55.96 / \textbf{65.32} & \textbf{65.59} / 56.60 / 65.23 & 65.50 / \textbf{58.05} / 65.27 & 65.46 / 58.00 / 65.10 \\
     & RML2022 & 62.91 / 56.25 / 66.03 & 64.28 / 56.43 / 66.11 & \textbf{64.36} / 57.73 / 66.84 & 63.70 / 58.01 / 66.58 & 63.25 / \textbf{58.35} / \textbf{67.01} \\
     & RML2018.01a & 60.21 / 51.63 / 63.97 & 64.58 / \textbf{56.96} / 64.17 & \textbf{64.70} / 56.54 / 64.32 & 64.19 / 56.33 / 64.22 & 64.54 / 56.53 / \textbf{64.70} \\
     & Sig2019-12 & 54.84 / 56.66 / 69.28 & 63.12 / 60.17 / 70.52 & 67.86 / \textbf{64.87} / \textbf{71.10} & \textbf{68.40} / \textbf{64.87} / 70.93 & 64.87 / \textbf{64.87} / 70.81 \\
     & HKDD\_AMC12 & 42.83 / 42.72 / 60.06 & 47.28 / 48.59 / 61.45 & 51.09 / \textbf{51.16} / 63.06 & 50.56 / \textbf{51.16} / \textbf{63.08} & \textbf{51.16} / \textbf{51.16} / 62.92 \\
     & HKDD\_AMC36 & 55.46 / 53.56 / 65.36 & 62.22 / \textbf{56.71} / 66.62 & \textbf{63.22} / 51.16 / \textbf{67.33} & 63.17 / 51.16 / 65.75 & 60.19 / 51.16 / 65.92 \\
     & HisarMod2019.1 & 54.42 / 51.79 / 62.63 & 63.72 / 54.32 / 74.10 & 78.11 / 78.11 / 78.30 & 80.69 / 80.69 / 80.53 & \textbf{82.30} / \textbf{82.30} / \textbf{80.90} \\
    \midrule
    \multirow{2}{*}{RWC} 
     & RadChar-Base & 89.70 / 84.72 / 90.08 & \textbf{89.72} / 84.80 / 90.06 & 89.55 / \textbf{89.55} / \textbf{90.13} & 89.50 / 89.50 / 90.06 & 89.34 / 89.34 / 90.12 \\
     & DeepRadar2022 & 66.86 / 66.62 / 77.45 & 77.82 / 69.26 / 83.90 & 76.07 / 76.07 / 87.07 & 74.99 / 74.99 / 86.61 & \textbf{78.06} / \textbf{78.06} / \textbf{87.45} \\
    \midrule
    \multirow{2}{*}{SEI} 
     & WiSig-ManyTX & 55.10 / 30.43 / 78.36 & 68.81 / 41.81 / 90.53 & 86.28 / 86.28 / 91.37 & 87.94 / 87.94 / \textbf{91.64} & \textbf{88.98} / \textbf{88.98} / 91.61 \\
     & ADSB-100 & 21.72 / 56.01 / 67.82 & 31.46 / 76.63 / 87.62 & 57.30 / 57.30 / 96.85 & 89.94 / 89.94 / 99.19 & \textbf{98.07} / \textbf{98.07} / \textbf{99.41} \\
    \midrule
    \multirow{2}{*}{WTR} 
     & SubGHz & 83.24 / 69.46 / 83.93 & 83.50 / 76.29 / 84.31 & \textbf{83.78} / \textbf{83.78} / 84.23 & 83.68 / 83.68 / \textbf{84.60} & 83.76 / 83.76 / 84.41 \\
     & TechRec & 77.94 / 79.83 / 88.72 & 81.26 / 80.11 / 89.15 & 84.39 / \textbf{82.58} / \textbf{89.70} & 84.74 / 82.53 / 89.64 & \textbf{85.00} / 82.27 / 89.44 \\
    \midrule
    WII 
     & EM-Infer-Comm & 80.98 / 74.84 / 83.08 & 82.16 / 75.13 / 83.75 & \textbf{82.83} / 75.29 / \textbf{84.02} & 82.49 / \textbf{75.71} / 83.93 & 81.99 / 75.46 / 83.72 \\
    \midrule
    Avg & All Tasks & 61.82 / 59.00 / 72.33 & 66.95 / 63.16 / 76.04 & 71.66 / 68.17 / 77.54 & 74.00 / 70.73 / 77.68 & \textbf{74.55} / \textbf{71.66} / \textbf{77.79} \\
    \bottomrule
    \end{tabular}
\end{table*}

\begin{figure}[htbp]
    \centering
    \includegraphics[width=0.95\linewidth]{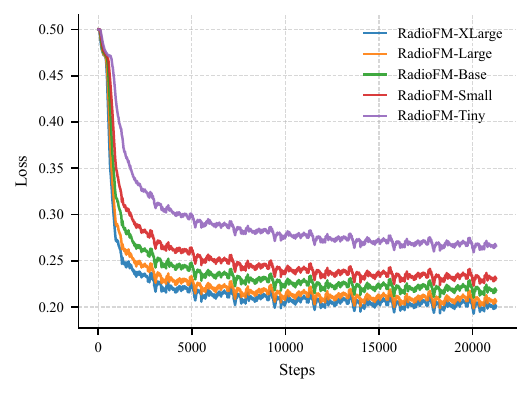}
    \caption{Pretraining Loss Convergence of Radio-FM Variants}	    
    \label{fig:scaling_loss}
\end{figure}

% \begin{figure}[htbp]
%     \centering
%     \includegraphics[width=1.00\linewidth]{Figures/model_scaling_paradigms.pdf}
%     \caption{Average accuracy across downstream tasks under different training paradigms for Radio-FM model scales. The bottom axis reports parameter counts, and the top axis reports FLOPs measured with signal length $N{=}1024$ and patch length/stride $8/8$.}
%     \label{fig:model_scaling_paradigms}
% \end{figure}

\begin{figure*}[htbp]
    \centering
    \begin{subfigure}[b]{0.32\textwidth}
        \centering
        \includegraphics[width=\textwidth]{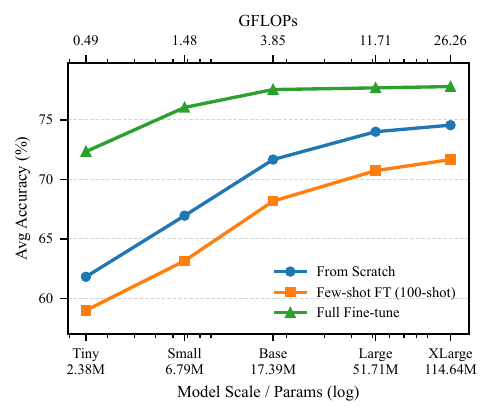}
        \caption{Model scaling paradigms}
        \label{fig:model_scaling_paradigms}
    \end{subfigure}
    \hfill
    \begin{subfigure}[b]{0.32\textwidth}
        \centering
        \includegraphics[width=\textwidth]{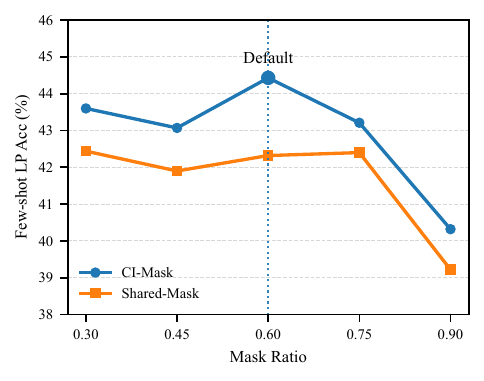}
        \caption{Mask ratio ablation}
        \label{fig:mask_ratio_ablation}
    \end{subfigure}
    \hfill
    \begin{subfigure}[b]{0.32\textwidth}
        \centering
        \includegraphics[width=\textwidth]{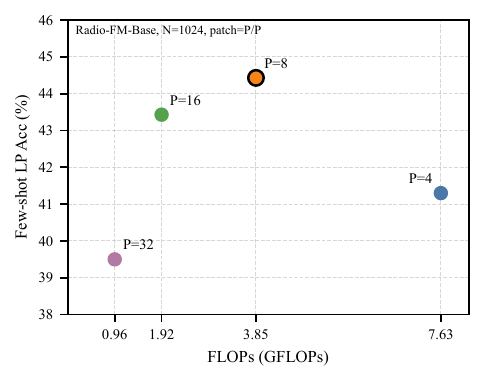}
        \caption{Patch size ablation}
        \label{fig:patch_size_ablation}
    \end{subfigure}
    \caption{Ablation Studies on Radio-FM Scaling and Design Choices. (a) Average accuracy across model scales under different training paradigms (parameter counts and FLOPs reported for $N{=}1024$, patch $8/8$). (b) Mask ratio comparison between CI-Mask and Shared-Mask under 100-shot linear probing. (c) Patch size trade-off between FLOPs and accuracy under 100-shot linear probing (Radio-FM-Base, $N{=}1024$).}	    \label{fig:ablation_studies}
\end{figure*}

We investigate the impact of model capacity on learning efficiency and transferability.
Fig.~\ref{fig:scaling_loss} reveals that increasing model size from Tiny to XLarge not only lowers the convergence loss but also significantly accelerates the learning rate, particularly in the initial 5k steps, indicating that larger models more efficiently compress the complex I/Q signal distributions.
Looking at downstream transfer in Fig.~\ref{fig:model_scaling_paradigms}, we observe distinct scaling behaviors across paradigms.
For \textit{Training from Scratch} and \textit{Few-shot FT}, performance exhibits a consistent log-linear growth with parameter count, confirming that larger capacities yield richer representations.
However, \textit{Full Fine-tuning} reveals a saturation effect: gains diminish notably beyond the Base scale (77.5\% vs. 77.8\% for XLarge).
This suggests that while larger models possess superior representational power (critical for data-scarce regimes), the performance ceiling in full-data regimes may be constrained by the intrinsic aleatoric uncertainty of the datasets rather than model capacity.

To further dissect these trends, Table~\ref{tab:consolidated_results} details the performance across all 15 datasets. A key observation is that the ``scaling dividend'' is non-uniform: high-complexity tasks like SEI (e.g., ADSB-100) and Radar (e.g., DeepRadar2022) witness massive gains from scaling (up to +76.35\% in ``From Scratch'' accuracy for ADSB-100 as model size grows), whereas simpler modulation tasks (e.g., RML2016 series) saturate quickly.
This indicates that Radio-FM's larger variants are particularly adept at capturing the fine-grained, localized features required for device fingerprinting and waveform characterization, which are often lost by smaller architectures.

This scaling benefit is visually corroborated by the prerained feature projections in the top rows of Fig.~\ref{fig:tsne_comparison}--Fig.~\ref{fig:tsne_techrec_comparison}.
At the Base scale (e.g., Fig.~\ref{fig:tsne_radchar_radiofm_base_pre}), the unsupervised representations show rudimentary clustering but significant scatter.
However, as we scale to Large and XLarge (Fig.~\ref{fig:tsne_radchar_radiofm_large_pre}--\ref{fig:tsne_radchar_radiofm_xlarge_pre}), the manifolds spontaneously organize into sharper, class-aligned structures despite the absence of supervision.
This distinct improvement in intrinsic feature quality elucidates why larger models exhibit superior few-shot transferability: they initialize fine-tuning from a highly structured latent space, reducing the need for extensive task-specific data.

\subsubsection{Mask Ratio Ablation}
Fig.~\ref{fig:mask_ratio_ablation} investigates the interplay between masking strategies and representation quality using 100-shot linear probing.
We identify an optimal mask ratio of 60\%, balancing task difficulty with contextual sufficiency. Notably, accuracy drops sharply at 90\%, indicating that excessive masking destroys the structural integrity of radio signals.
Crucially, the Channel-Independent (CI) Mask consistently outperforms the Shared Mask strategy by a margin of $\sim$2\%, with the gap widening at the optimal ratio.
This confirms that uncoupling the mask patterns forces the model to leverage cross-channel correlations---reconstructing the missing I-component from the visible Q-component (and vice versa)---thereby learning more robust, phase-aware representations.

\subsubsection{Patch Size Ablation}
% \begin{figure}[htbp]
%     \centering
%     \includegraphics[width=1.00\linewidth]{Figures/patch_size_ablation.pdf}
%     \caption{Patch size ablation under 100-shot linear probing (Radio-FM-Base). We report the FLOPs--accuracy trade-off, where FLOPs are evaluated with signal length $N{=}1024$ and patch length/stride $P/P$.}
%     \label{fig:patch_size_ablation}
% \end{figure}

Fig.~\ref{fig:patch_size_ablation} characterizes the critical trade-off between temporal resolution and computational efficiency.
We identify a distinct peak at $P=8$, providing an optimal balance (44.4\% accuracy at 3.85 GFLOPs).
Contrary to the intuition that finer granularity improves performance, reducing the patch size to $P=4$ degrades accuracy to 41.3\%. Although this incurs a $2\times$ FLOPs penalty, the drop suggests that over-fragmenting the signal disrupts local semantic coherence, complicating the extraction of meaningful global patterns.
Conversely, increasing $P$ to 32 leads to a sharp performance decline (39.5\%), as coarse-grained patching fails to resolve the transient micro-features essential for modulation and emitter identification.

\subsubsection{Module Ablation}
Table~\ref{tab:ablation_module} quantitatively validates the contribution of each design component using the Radio-FM-Base backbone.
Reverting to a shared masking strategy incurs the largest performance penalty (-2.11\%), confirming that channel-independent masking imposes a harder pretext task that forces the model to learn robust inter-channel correlations rather than simple interpolation.
Regarding architectural components, completely removing the dual-channel interaction module (`w/o Inter-channel`) degrades accuracy by 0.81\%, validating the necessity of explicit I/Q information exchange.
Furthermore, simplifying this interaction by removing its layer scaling (`w/o LayerScale`) leads to a 0.57\% drop. This suggests that the dynamic recalibration of I/Q features is essential for capturing the non-stationary phase variations inherent in radio signals.

\begin{table}[htbp]
    \centering
	\caption{Ablation Study on Channel Processing Components and Masking Strategies. Validation of the effectiveness of inter-channel interaction, gating mechanism, and Channel-Independent Mask (CI-Mask).}    \label{tab:ablation_module}
    \begin{tabular}{l c c c c}
    \toprule
    \multirow{2}{*}[-1ex]{\textbf{Method}} & \multicolumn{3}{c}{\textbf{Module}} & \multirow{2}{*}[-1ex]{\textbf{Acc (\%)}} \\
    \cmidrule(lr){2-4}
     & \makecell{\textbf{Inter-}\\\textbf{Channel}} & \makecell{\textbf{LayerScale}} & \makecell{\textbf{CI-}\\\textbf{Mask}} & \\
    \midrule
    w/o Inter-channel       &            & \checkmark & \checkmark & 43.62 \\
    w/o LayerScale          & \checkmark &            & \checkmark & 43.86 \\
    w/o CI-Mask             & \checkmark & \checkmark &            & 42.32 \\
    \textbf{Radio-FM-Base}  & \checkmark & \checkmark & \checkmark & \textbf{44.43} \\
    \bottomrule
    \end{tabular}
\end{table}

\subsection{Discussion}
While Radio-FM demonstrates promising results, several challenges and opportunities persist for the advancement of radio foundation models.
\textbf{Data Standardization and Scale.} Unlike Computer Vision (CV) and Natural Language Processing (NLP), which benefit from massive, unified corpora (e.g., ImageNet, CommonCrawl), the RF domain is characterized by extreme fragmentation.
Signal collection, simulation standards, and storage formats vary significantly across research groups. In this study, aligning benchmarks across 15 datasets required substantial data cleaning and standardization.
Establishing a unified, large-scale open-source corpus with consistent preprocessing protocols is a critical prerequisite for training next-generation models with broader generalization capabilities.
\textbf{Multimodal Representation Learning.} Radio-FM currently primarily utilizes raw I/Q sequences, which serve as the fundamental representation of the physical layer.
However, communication systems can be analyzed via multiple modalities, such as time-frequency spectrograms, constellation diagrams, and Channel State Information (CSI).
Future architectures could benefit from multimodal fusion, integrating these complementary perspectives to construct a more holistic representation of the electromagnetic environment.
\textbf{Synergy with Large Language Models (LLMs).} Bridging embedding-based radio models with the reasoning capabilities of LLMs constitutes an emerging research frontier.
By aligning the perceptual capabilities of Radio-FM with the reasoning mechanisms of LLMs, we envision the development of ``Electromagnetic Agents'' capable of semantic communications, automated protocol optimization, and interpretable spectrum reasoning, thereby transcending simple pattern recognition.
\textbf{Edge Adaptation and Efficiency.} Our scaling analysis confirms that increasing the number of parameters yields superior transferability.
However, deploying such billion-parameter backbones on resource-constrained edge devices (e.g., IoT nodes, drones) remains a significant challenge.
Future research must address this gap through lightweight techniques such as knowledge distillation, quantization, and structural pruning, ensuring that the benefits of large-scale pretraining are accessible in latency-sensitive, low-power applications.
\section{Conclusion\label{sec_4}}
In this paper, we presented Radio-FM, a scalable foundation model framework designed for universal radio signal representation learning. 
Addressing the unique characteristics of RF data, we proposed a dual-channel processing architecture specifically optimized for I/Q independence, coupled with a robust pretraining recipe that handles extreme data heterogeneity via channel-independent masked reconstruction and dynamic batching. 
Through rigorous evaluation on 15 datasets spanning modulation, radar, and communication domains, Radio-FM demonstrated state-of-the-art performance on 13 benchmarks and achieved significant gains in few-shot data efficiency compared to existing baselines. 
Our findings confirm that scaling model capacity on diverse raw I/Q corpora yields emergent capabilities crucial for complex signal understanding. 
We envision Radio-FM as a pivotal step towards general-purpose electromagnetic intelligence, paving the way for future advancements in multimodal integration and autonomous spectrum agents.

% \appendix
% \section{Appendix A: Pretraining Loss}
% \section{Appendix B: Downstream Task Performance}
% \section{Appendix C: Model Variants}
% \section{Appendix D: Comparison to State of the Art}

\bibliographystyle{ieeetr}
\bibliography{REFs}

\end{document}